\documentclass[aps,prd,reprint,superscriptaddress,floatfix,nofootinbib,showkeys,showpacs,preprintnumbers]{revtex4-2}
\usepackage{tabularx}
\usepackage{xspace}
\usepackage{listings}
\usepackage{lineno}
\usepackage[T1]{fontenc}

\usepackage{bm}		%
\usepackage{amsmath}	%
\usepackage{amssymb}	%
\usepackage[normalem]{ulem}

\usepackage[dvipsnames,svgnames]{xcolor} 
\usepackage{graphicx}
\usepackage{comment}

\usepackage{natbib}
\usepackage{hyperref}
\usepackage{amsmath}
\usepackage{enumitem}
\setlist{wide, labelwidth=!, labelindent=0pt}
\usepackage{multirow}
\hypersetup{    
  colorlinks      = true,
  linkcolor       = {black},
  linkbordercolor = {white},
  citecolor       = {black},
  citebordercolor = {white},
  urlcolor        = {blue},
  urlbordercolor  = {white},
}
\usepackage{soul} 
\usepackage{amsmath}
\usepackage{amssymb}
\usepackage{xspace}
\usepackage{xifthen}

\mathchardef\mhyphen="2D

\newlength{\dhatheight}

\newcommand{\unit}[1]{\ensuremath{\mathrm{\,#1}}\xspace}

\newcommand{\Msolar}{\unit{M_\odot}}
\newcommand{\Msun}{\unit{M_\odot}}

\newcommand{\bandvar}[2][]{%
  \ifthenelse{\isempty{#1}}{\var{#2}}{\var{#2\_#1}}%
}

\newcommand{\LCDM}{\ensuremath{\rm \Lambda CDM}\xspace}

\newcommand{\var}[1]{\ensuremath{\texttt{\MakeUppercase{#1}}}\xspace}

\defcitealias{Simpaper}{Simpaper}
\defcitealias{DESY1KP}{DESY1KP}
\defcitealias{DES_cluster_cosmology}{DESCL}

\providecommand\physrep{\ref@jnl{Phys.~Rep.}}%
\providecommand\apjs{\ref@jnl{ApJS}}%
\providecommand{\jcap}{\ref@jnl{JCAP}}%

\newcommand{\RNum}[1]{\uppercase\expandafter{\romannumeral #1\relax}}

 \usepackage{lipsum}
\usepackage{aas_macros}

\usepackage{tikz}
\usepackage{calc}

\begin{document}

\title{ A DECADE of dwarfs: first detection of weak lensing around spectroscopically confirmed low-mass galaxies}

\author{
Chun-Hao To} \thanks{Corresponding author: \href{mailto:chto@uchicago.edu}{chto@uchicago.edu}}
\affiliation{Department of Astronomy and Astrophysics, University of Chicago, Chicago, IL 60637, USA}
\affiliation{Kavli Institute for Cosmological Physics, University of Chicago, Chicago, IL 60637, USA}
\affiliation{NSF-Simons AI Institute for the Sky (SkAI),172 E. Chestnut St., Chicago, IL 60611, USA}
\author{
Chihway Chang}
\affiliation{Department of Astronomy and Astrophysics, University of Chicago, Chicago, IL 60637, USA}
\affiliation{Kavli Institute for Cosmological Physics, University of Chicago, Chicago, IL 60637, USA}
\affiliation{NSF-Simons AI Institute for the Sky (SkAI),172 E. Chestnut St., Chicago, IL 60611, USA}
\author{
Dhayaa Anbajagane}
\affiliation{Department of Astronomy and Astrophysics, University of Chicago, Chicago, IL 60637, USA}
\affiliation{Kavli Institute for Cosmological Physics, University of Chicago, Chicago, IL 60637, USA}
\affiliation{NSF-Simons AI Institute for the Sky (SkAI),172 E. Chestnut St., Chicago, IL 60611, USA}
\author{
Risa H. Wechsler}
\affiliation{Department of Physics, Stanford University, 382 Via Pueblo Mall, Stanford, CA 94305, USA}
\affiliation{Kavli Institute for Particle Astrophysics \& Cosmology, P. O. Box 2450, Stanford University, Stanford, CA 94305, USA}
\affiliation{SLAC National Accelerator Laboratory, Menlo Park, CA 94025, USA}
\author{
Alex Drlica-Wagner}
\affiliation{Department of Astronomy and Astrophysics, University of Chicago, Chicago, IL 60637, USA}
\affiliation{Kavli Institute for Cosmological Physics, University of Chicago, Chicago, IL 60637, USA}
\affiliation{NSF-Simons AI Institute for the Sky (SkAI),172 E. Chestnut St., Chicago, IL 60611, USA}
\affiliation{Fermi National Accelerator Laboratory, P.O.\ Box 500, Batavia, IL 60510, USA}

\author{M.~Adam\'ow}
\affiliation{Center for Astrophysical Surveys, National Center for Supercomputing Applications, 1205 West Clark St., Urbana, IL 61801, USA}
\affiliation{Department of Astronomy, University of Illinois at Urbana-Champaign, 1002 W. Green Street, Urbana, IL 61801, USA}

\author{A.~Alarcon}
\affiliation{Institute of Space Sciences (ICE, CSIC),  Campus UAB, Carrer de Can Magrans, s/n,  08193 Barcelona, Spain}

\author{M.~R.~Becker}
\affiliation{Argonne National Laboratory, 9700 South Cass Avenue, Lemont, IL 60439, USA}

\author{J. A. Carballo-Bello}
\affiliation{Instituto de Alta Investigacion, Universidad de Tarapaca, Casilla 7D, Arica, Chile}

\author{R.~Cawthon}
\affiliation{Oxford College of Emory University, Oxford, GA 30054, USA}

\author{N.~Chicoine}
\affiliation{Department of Astronomy and Astrophysics, University of Chicago, Chicago, IL 60637, USA}
\affiliation{Department of Physics and Astronomy, University of Pittsburgh, 3941 O’Hara Street, Pittsburgh, PA 15260}

\author{C.~Doux}
\affiliation{Université Grenoble Alpes, CNRS, LPSC-IN2P3, 38000 Grenoble, France}

\author{J.~H.~Esteves}
\affiliation{Department of Physics, Harvard University, 17 Oxford St, Cambridge, MA, USA}

\author{P.~S.~Ferguson}
\affiliation{DiRAC Institute, Department of Astronomy, University of Washington, 3910 15th Ave NE, Seattle, WA, 98195, USA}
\affiliation{DIRAC Institute, Department of Astronomy, University of Washington, 3910 15th Ave NE, Seattle, WA, 98195, USA}

\author{M.~Gatti}
\affiliation{Kavli Institute for Cosmological Physics, University of Chicago,
Chicago, IL 60637, USA}

\author{D.~Gruen}
\affiliation{University Observatory, Faculty of Physics, Ludwig-Maximilians-Universität München, Scheinerstr. 1, 81679 Munich, Germany; Excellence Cluster ORIGINS, Boltzmannstr. 2, 85748 Garching, Germany}

\author{R.~A.~Gruendl}
\affiliation{Department of Astronomy, University of Illinois at Urbana-Champaign, 1002 W. Green Street, Urbana, IL 61801, USA }
\affiliation{Center for Astrophysical Surveys, National Center for Supercomputing Applications, 1205 West Clark St, Urbana, IL 61801, USA}
\affiliation{Center for Astrophysical Surveys, National Center for Supercomputing Applications, 1205 West Clark St., Urbana, IL 61801, USA}
\affiliation{Department of Astronomy, University of Illinois at Urbana-Champaign, 1002 W. Green Street, Urbana, IL 61801, USA}

\author{K.~Herron}
\affiliation{Department of Physics and Astronomy, Dartmouth College, Hanover, NH 03755, USA}

\author{David~J.~James}
\affiliation{ASTRAVEO LLC, PO Box 1668, Gloucester, MA 01931, USA}
\affiliation{Applied Materials Inc., 35 Dory Road, Gloucester, MA 01930, USA}

\author{C.~E.~Mart\'inez-V\'azquez}
\affiliation{NSF NOIRLab, 670 N. A'ohoku Place, Hilo, Hawai'i, 96720, USA}

\author{S.~Mau}
\affiliation{Department of Physics, Stanford University, 382 Via Pueblo Mall, Stanford, CA 94305, USA}
\affiliation{Kavli Institute for Particle Astrophysics \& Cosmology, P.O.\ Box 2450, Stanford University, Stanford, CA 94305, USA}

\author{J.~McCullough}
\affiliation{Department of Astrophysical Sciences, Princeton University, Peyton Hall, Princeton, NJ 08544, USA}

\author{G.~E.~Medina}
\affiliation{David A. Dunlap Department of Astronomy \& Astrophysics, University of Toronto, 50 St George Street, Toronto ON M5S 3H4, Canada}
\affiliation{Department of Astronomy and Astrophysics, University of Toronto, 50 St. George Street, Toronto ON, M5S 3H4, Canada}

\author{B.~Mutlu-Pakdil}
\affiliation{Department of Physics and Astronomy, Dartmouth College, Hanover, NH 03755, USA}

\author{A. ~Navarro-Alsina}
\affiliation{Instituto de Física Gleb Wataghin, Universidade Estadual de Campinas, 13083-859, Campinas, SP, Brazil}

\author{N.~E.~D. No\"el}
\affiliation{Department of Physics, University of Surrey, Guildford GU2 7XH, UK}

\author{A.~B.~Pace}
\affiliation{Department of Astronomy, University of Virginia, 530 McCormick Road, Charlottesville, VA 22904, USA}

\author{M.~Raveri}
\affiliation{Department of Physics and INFN, University of Genova, Via Dodecaneso 33, 16146, Italy}

\author{A.~H.~Riley}
\affiliation{Institute for Computational Cosmology, Department of Physics, Durham University, South Road, Durham DH1 3LE, UK}
\affiliation{Lund Observatory, Division of Astrophysics, Department of Physics, Lund University, SE-221 00 Lund, Sweden}

\author{D.~J.~Sand}
\affiliation{Steward Observatory, University of Arizona, 933 North Cherry Avenue, Tucson, AZ 85721-0065, USA}

\author{L.~F.~Secco}
\affiliation{Kavli Institute for Cosmological Physics, University of Chicago, Chicago, IL 60637, USA}

\author{T.~Shin}
\affiliation{Department of Physics and Astronomy, Carnegie Mellon University, Pittsburgh, PA 15213, USA}

\author{G.~S.~Stringfellow}
\affiliation{University of Colorado Boulder, Boulder, CO 80309, USA}

\author{D.~Suson}
\affiliation{Department of Chemistry and Physics, Purdue University Northwest 2200, 169th Ave, Hammond, IN 46323}

\author{C.~Y.~Tan}
\affiliation{Department of Physics, University of Chicago, Chicago, IL 60637, USA}
\affiliation{Kavli Institute for Cosmological Physics, University of Chicago, Chicago, IL 60637, USA}

\author{R.~Teixeira}
\affiliation{Department of Astronomy and Astrophysics, University of Chicago, Chicago, IL 60637, USA}
\affiliation{Department of Physics, Duke University Durham, NC 27708, USA}

\author{A.~Zenteno}
\affiliation{Cerro Tololo Inter-American Observatory/NSF NOIRLab, Casilla 603, La Serena, Chile}

\author{Z.~Zhang}
\affiliation{Department of Astronomy and Astrophysics, University of Chicago, Chicago, IL 60637, USA}
\affiliation{Department of Physics, Stanford University, 382 Via Pueblo Mall, Stanford, CA 94305, USA}
\affiliation{SLAC National Accelerator Laboratory, Menlo Park, CA 94025, USA}

\collaboration{DELVE Collaboration}

\begin{abstract}
We present the first detection of weak gravitational lensing around spectroscopically confirmed dwarf galaxies, using the large overlap between DESI DR1 spectroscopic data and DECADE/DES weak lensing catalogs. A clean dwarf galaxy sample with well-defined redshift and stellar mass cuts enables excess surface mass density measurements in two stellar mass bins ($\log \rm{M}_*=[8.2, 9.2]~\Msun$ and $\log \rm{M}_*=[9.2, 10.2]~\Msun$), with signal-to-noise ratios of $5.6$ and $12.4$ respectively. This signal-to-noise drops to $4.5$ and $9.2$ respectively for measurements without applying individual inverse probability (IIP) weights, which mitigates fiber incompleteness from DESI's targeting. The measurements are robust against variations in stellar mass estimates, photometric shredding, and lensing calibration systematics. Using a simulation-based modeling framework with stellar mass function priors, we constrain the stellar mass–halo mass relation and find a satellite fraction of $\simeq 0.3$, which is higher than previous photometric studies but $1.5\sigma$ lower than $\LCDM$ predictions. We find that IIP weights have a significant impact on lensing measurements and can change the inferred $f_{\rm sat}$ by a factor of two, highlighting the need for accurate fiber incompleteness corrections for dwarf galaxy samples. Our results open a new observational window into the galaxy–halo connection at low masses, showing that future massively multiplexed spectroscopic observations and weak lensing data will enable stringent tests of galaxy formation models and $\LCDM$ predictions.

\end{abstract}
\maketitle 

\section{Introduction}

Dwarf galaxies\footnote{Here, we follow the definition in \citep{Bullock17} and refer to dwarf galaxies as galaxies with stellar mass ($M_*$) less than $10^9~\Msun$.} are a unique probe of the nature of dark matter and galaxy formation models \citep[see e.g.,][for a review]{Bullock17}. Their matter profiles are sensitive probes of dark matter physics. Their star formation is sensitive to feedback from ionizing radiation, such as stellar winds and supernovae, due to the shallow gravitational potential of the host dark matter halos \citep{2022NatAs...6..647C,2002MNRAS.333..177B,2000ApJ...539..517B,2021MNRAS.508.2979P}. 

One way to study astrophysics with dwarf galaxies is to quantify the statistical relation of their stellar mass and the mass of the host dark matter halos, known as the galaxy--halo connection \citep{risaawesomepaper}\footnote{In general, the galaxy--halo connection describes the statistical relations of properties of galaxies and properties of host dark matter halos. In this paper, we refer to the stellar mass -- halo mass relation.}. The slope, normalization, and scatter of this relation has implications for a wide range of galaxy formation physics, including feedback-driven outflows \citep{1986ApJ...303...39D} and heating of intergalactic gas by UV photons during reionization \citep{2000ApJ...539..517B} (see also \citep{Grumpy} for pedalogical examples of the interplay between these effects). The galaxy--halo connection is also a key modeling ingredient for inferring dark matter physics from dwarf galaxy abundances \citep{Nadler21, Bullock17}. 

However, measuring the stellar mass--halo mass (SMHM) relations of dwarf galaxies is challenging. Numerous attempts have relied on relating galaxy abundances to halo abundances and a monotonic relation of galaxy stellar mass and halo mass, an approach known as abundance matching \citep{Nadler21, Behroozi13,Moster13}. While powerful, such an approach relies on assumptions about the stochasticity of stellar mass given halo mass, which is uncertain \citep{risaawesomepaper,2024arXiv240815214K}. A more direct approach is to measure the halo mass through stellar and gas kinematics \citep{1998ApJ...499...41M,2019MNRAS.484.1401R,2019MNRAS.487.5799R}. However, kinematic measurements usually only extend out to a few kpc, much smaller than the virial radius of the dark matter halos. Thus, it is usually necessary to assume a specific dark matter halo density profile in order to infer the total halo mass from such a measurement.

Weak gravitational lensing provides a promising alternative to accurately determine the mass of dark matter halos. It measures the coherent distortions of the background galaxy shapes around the lens galaxies and is sensitive to matter distributions around dark matter halos across a wide range of scales. Inferring total halo mass from such measurements requires fewer assumptions about the shape and properties of dark matter halos. However, weak gravitational lensing is weak ($\rm{S/N} \simeq1\%$ for each galaxy), requiring a large number of pairs of foreground ``lens'' and background ``source'' galaxies.

Detecting a large number of extragalactic dwarf galaxies over a wide area of sky is challenging. They are typically only detectable in current wide-area imaging surveys at low redshift. %
Separating dwarf galaxies from the dominant background of high-redshift galaxies is challenging \citep{saga1, 2023ApJ...954..149D, SAGA2, saga3}. As a result, existing spectroscopically confirmed dwarfs \citep{saga3} have historically been too sparse to yield a detection of gravitational lensing. To mitigate this problem, \citep{Thornton,2024OJAp....7E..89C} relied on photometrically selected dwarf galaxies. The authors built a relation between galaxy stellar mass and redshift distributions and their broadband photometries using spectroscopically selected dwarf samples. They then measure weak lensing profiles around photometrically selected dwarf galaxies. While this approach yields high signal-to-noise detections of lensing profiles, the redshift uncertainties of the samples are usually large, so that the mean redshift distribution of the photometrically selected samples differs from the spectroscopically selected samples. Since several galaxy properties, such as colors, evolve as a function of redshifts, this mismatch can make it challenging to quantify the selection functions of the samples, complicating the interpretation of the measurements \citep{Thornton,2024OJAp....7E..89C}.

In this work, we leverage the significant overlap between a large spectroscopic survey (Dark Energy Spectroscopic Survey or DESI \citep{2025arXiv250314745D}) and a photometric weak lensing survey (the Dark Energy Camera All Data Everywhere or DECADE \citep{Anbajagane2025a}) to measure weak gravitational lensing profiles around spectroscopically selected dwarf galaxies. DESI has dramatically increased the number of spectroscopically confirmed dwarf galaxies through its Bright Galaxy Survey \citep{BGS} and its LOW-Z Secondary Target Survey \citep{Elise}. Through a combination of these and other programs, the DESI first data release has provided a sample of $>350$k dwarf galaxies ($\log (M_*/\Msun)<9$), tripling those from Sloan Digital Sky Survey (SDSS) \citep{SDSS} and Galaxy And Mass Assembly Survey \citep{GAMA}. On the lensing side, the DECADE weak lensing data has similar qualities as the Dark Energy Survey Year 3 data (DES-Y3, \citep{y3-shapecatalog})  and has almost tripled the overlap between Stage-3 quality weak lensing data (i.e, the combined overlap of the weak lensing catalogs from the Dark Energy Survey \citep{DES}, the Hyper Suprime-Cam Subaru Strategic Program \citep{2022PASJ...74..421L}, and the Kilo Degree Survey \citep{2021A&A...645A.105G}) and the DESI spectroscopic samples. With DECADE and DESI, we present the first detection of lensing profiles around spectroscopically confirmed dwarf galaxies. With these measurements, we employ a simulation-based inference pipeline to extract the galaxy stellar mass--host halo mass relation and the satellite fraction of the sample. Our samples have a relatively simple selection function compared to photometrically selected samples \citep{Thornton, 2024OJAp....7E..89C}, making it relatively straightforward to interpret these findings. 
\begin{figure*}
\includegraphics[width=0.9\linewidth]{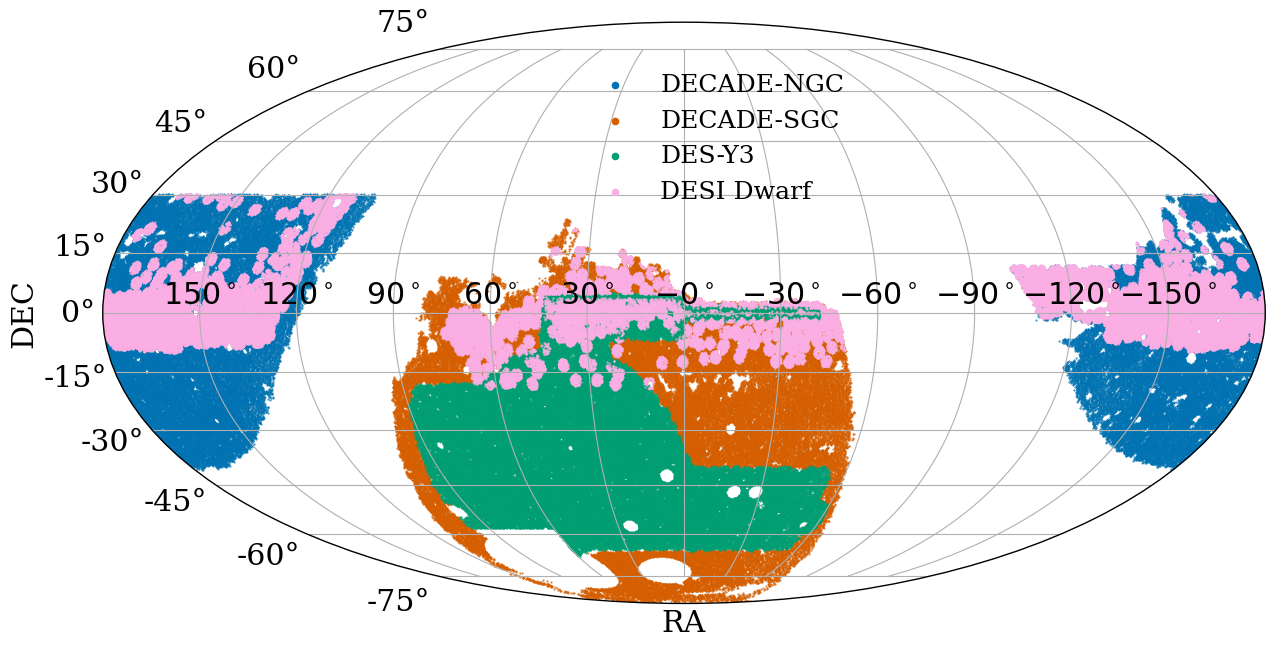}
\caption{The footprint of datasets used in this analysis: DECADE-NGC (blue), DECADE-SGC (orange), DES-Y3 (green), and DESI Dwarf (pink), shown in the Mollweide projection. The decade data is split into the North Galactic Cap (NGC) and South Galactic Cap (SGC) regions, as described in section \ref{sec:decade}. The fraction of overlaps is summarized in Table \ref{tab:numgal}.}
\label{fig:lensing}

\end{figure*}

This paper is organized as follows. In Section \ref{sec:Data}, we detail the construction of the dwarf galaxy samples and the lensing sample. Section \ref{sec:measurementandmodeling} provides a detailed description of the weak lensing measurements and our simulation-based model. Section \ref{sec:results} shows the constrained stellar mass--halo mass relation and the satellite fraction. We conclude in Section \ref{sec:conc}. Throughout the paper, $\log$ denotes $\log_{10}$.

\section{Data}
\label{sec:Data}

We utilize the overlapped area between the Dark Energy Spectroscopic Instrument Data Release 1 (DESI-DR1) Bright Galaxy Survey (BGS) and the shear catalogs from the DECam All Data Everywhere (DECADE) as well as the Dark Energy Survey Year 3 (DES-Y3) dataset. Figure~\ref{fig:lensing} shows the footprint of each of the surveys as well as the overlap. The overlap region is about $4303\ \rm{deg}^2$ in total.

\subsection{DESI BGS dwarf sample}
In this paper, we use the Extragalactic Dwarf Catalog\footnote{\url{https://data.desi.lbl.gov/doc/releases/dr1/vac/extragalactic-dwarfs/}; Manwadkar+ in preparation}, a compilation of low-stellar-mass ($M_* (\rm{CIGALE})<10^{10} \Msun$) galaxies identified across multiple DESI-DR1 programs, including BGS, emission-line galaxy samples, and the LOW-Z Secondary Target Program \citep{dwarfcatalog}. Due to this multi-program origin, characterizing the overall selection function for the full sample is nontrivial. To simplify the analysis and to ensure a well-defined selection function, we focus on the ``Bright'' subset of the BGS sample (BGS-Bright), which is selected with $r$-band magnitude ${m}_{r}$ brighter than $19.5$ \citep{BGS}. We note that the BGS-Bright sample has the highest priority of fiber assignment in the BGS program. Potential galaxies in the BGS-Bright samples are assigned a fiber with the same priority regardless of their properties, such as flux and colors. During the survey, the only target that has a higher priority than BGS-Bright is DESI Milky Way Survey's white dwarf sample \citep{2023ApJ...947...37C}, whose number density is $~\sim 4.3$ per $\rm{deg}^2$, $180$ times smaller than BGS-Bright targets. Below, we outline and justify our analysis choices, including methods for stellar mass estimation, binning in redshift and stellar mass, and the treatment of implicit selection effects.

\subsubsection{Stellar mass estimation}
\label{sec:stellarmass}
A number of stellar mass estimations have been provided by the DESI collaboration \footnote{We note that the photometry used in these stellar mass estimates is from Tractor, which can be unreliable for shredded systems. \citep{dwarfcatalog} is developing a new pipeline for shredded systems. The  \textsc{$MAG$\_\{G,R,Z\}\_UPDATED} in the public DESI dwarf catalog provides a combination of Tractor photometry and photometry from the new pipeline, using the best estimate from the authors of \citep{dwarfcatalog} on which measurement is most reliable for each source.  Since the criteria for selecting between the two is \textsc{fracflux}$>0.35$, we are already testing that with \textsc{fracflux}$=0.2$ cut, these differences in photometry do not impact our analysis.}, including: 
\begin{enumerate}
    \item SAGA: a color-based stellar mass using $r$-band absolute magnitude and restframe $g-r$ color \citep{SAGA2}. The absolute magnitude and colors are $k$-corrected to $z=0$ \citep{2010MNRAS.405.1409C}.
    \item CIGALE \citep{2019A&A...622A.103B}: an SED-based stellar mass estimation using optical $g,r,z$ and infrared bands $W_1, W_2, W_3, W_4$ \citep{CIGALE}. In the implementation of \citep{CIGALE}, the metallicity is fixed to the solar value, and AGN templates have been considered in the SED fitting procedure. 
    \item CIGALE-noAGN \citep{CIGALE2}: the input information is the same as \citep{CIGALE}, but no AGN templates are considered, and the metallicity is varied. 
\end{enumerate}

Appendix~\ref{sec:stellar_mass} compares the stellar masses obtained from these three methods. We find good agreement between SAGA and CIGALE-noAGN, while CIGALE and CIGALE-noAGN differ by approximately $0.15$ dex. Since the dwarf galaxies considered in this work have stellar masses significantly below that of the Milky Way, we expect their metallicities to also be below the solar value. Further, as shown in appendix \ref{sec:stellar_mass}, the contribution of AGN to the total infrared light in CIGALE estimation seems to be overestimated due to the low signal-to-noise of the infrared data. We therefore adopt the CIGALE-noAGN estimates as our fiducial stellar masses, and defer a detailed investigation of the offset between CIGALE and CIGALE-noAGN to future work.  The agreement between the SED-based (CIGALE-noAGN) and the simpler photometric method (SAGA) further supports this choice. Given that assumptions in SED modeling can introduce systematic biases in stellar mass estimates \citep{2024arXiv240903959D}, the consistency across these independent approaches increases our confidence in the robustness of the adopted masses. In section \ref{sec:mr}, we further investigate the differences in the data vector with different choices of stellar mass estimates.

\subsubsection{Redshift and stellar mass binning}

\begin{figure}
\includegraphics[width=\linewidth]{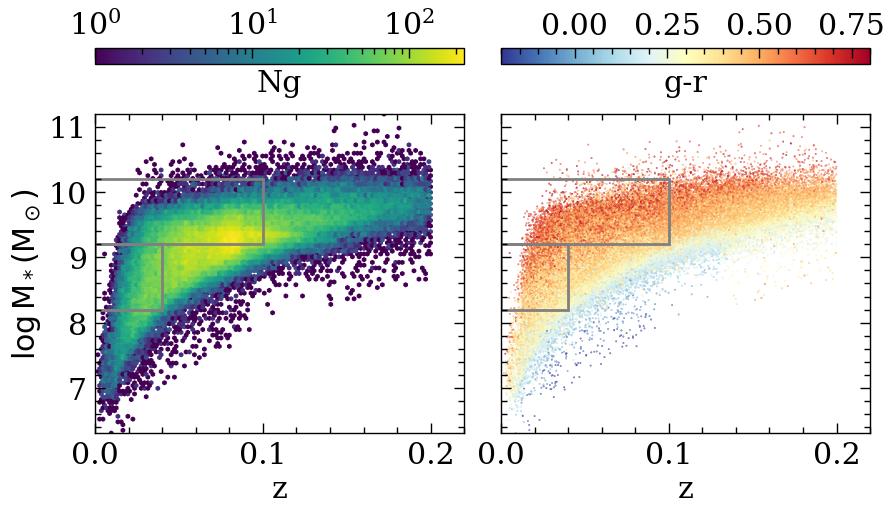}
\caption{Distributions of the galaxy number (left) and rest-frame $g-r$ colors (right) of the DESI BGS Bright sample in the CIGALE-noAGN stellar mass vs. redshift plane. The gray boxes indicate the samples used in this analysis.}
\label{fig:sampleselect}
\end{figure}
Figure~\ref{fig:sampleselect} shows the redshift and stellar mass distribution of the dwarf galaxy sample. As expected near the detection limit, the selected galaxies are predominantly blue, consistent with findings from SDSS DR7 \citep{ZuandMandelbaum}. This bias arises because, at fixed stellar mass, star-forming galaxies are typically brighter than quiescent ones, making them more likely to be selected in a flux-limited sample. To avoid complications in modeling this color-dependent selection effect, we remove blue galaxies near the detection limits to mitigate biasing the target sample. As a result, we restrict our analysis to the region indicated by the gray boxes in Fig.~\ref{fig:sampleselect}. 
To quantitatively evaluate the validity of the redshift cut, we compare the median $g-r$ colors of galaxies with redshifts below the redshift cut to those of galaxies within $0.01$ in redshift of the cut. This comparison is performed at the low-mass boundary of each stellar mass bin. Specifically, we define
\begin{eqnarray} &&D_{x,y}=\nonumber\\ &&\rm{Median} (g-r ([z \in (x-0.01, x), \log (\frac{M_*}{\Msun})\in (y, y+0.01)]) \nonumber\\ &&-\rm{Median} (g-r([z<x, \log(\frac{M_*}{\Msun}) \in (y, y+0.01)]) ,\end{eqnarray}
where $x$ is the redshift cut and $y$ is the stellar mass boundary. We evaluate this at $(x,y) = (0.04, 8.2)$ and $(0.1, 9.2)$. By construction, $D_{x,y}=0$ if galaxies near the redshift cut have similar median colors to those well within the cut, while negative values indicate that galaxies near the cut are bluer. We require $D_{x,y}>-0.05$ as our criterion for a valid cut. At $\log (M_*/\Msun)=8.2$ and $9.2$, we find $D_{x,y}=-0.02$ and $-0.04$, respectively. Increasing the redshift cut by $0.01$ yields larger deviations of $-0.10$ and $-0.07$.

\subsubsection{Implicit target selections}
Not all galaxies with $m_r<19.5$ are observed by DESI due to the limiting number of fibers on the focal plane. This limitation means that DESI spectroscopic samples tend to undersample overdense regions on the sky \citep{BGS_fiber_09}. Furthermore, extended objects will have less flux within the fibers, making their redshift failure rates higher. These implicit target selections correlate with galaxy size and large-scale environments, thus biasing the galaxy--galaxy lensing signals. 

To mitigate the problem of fiber incompleteness, we use the individual inverse probability, or IIP, weight ($w_{\rm{IIP}}$) from DESI, which has been shown on Buzzard mocks \citep{Buzzard} to successfully mitigate this implicit target selection problem \citep{desilensing}. The IIP weight is constructed by Monte Carlo sampling DESI's fiber-assignment algorithms on photometrically selected potential targets. Specifically, DESI records, for each observation, the number of objects that would have been observed by each fiber \citep{DESItarget,2025JCAP...07..017A}. Since only one of these potential targets will be observed, the inverse of this number (IIP weight) mitigates the incompleteness of the fiber assignment due to the competition of the fibers. We note that the IIP weight only depends on the priority with respect to fiber assignments. Since all potential BGS-Bright targets have the same priority, the validity of the IIP weight does not depend on the magnitude and redshift cuts to select dwarf galaxies. We further corroborate this argument by performing a dedicated simulation test in appendix \ref{sec:fiberassignment}.

Not every galaxy with a fiber assigned will have a successful redshift estimate. Since we only use galaxies that have a successful redshift estimate, we need to mitigate this implicit selection effect. Here, we use the redshift failure weight ($w_{\rm{zfail}}$) provided by DESI. Redshift failure is expected to correlate with the total flux observed by the fiber and the spectrum signal-to-noise ratio \citep{2025JCAP...01..147K}. DESI empirically constructs the redshift failure rate model based on the fiber flux and the signal-to-noise ratio \citep{DESItarget,2025JCAP...01..147K}. The redshift failure weight ($w_{\rm{zfail}}$)  is then determined by the inverse of the predicted redshift failure rate.

In the main analysis, we weigh each galaxy by the product of  $w_{\rm{zfail}}$ and $w_{\rm{IIP}}$. We discuss the impact of these weights in our analysis in section \ref{sec:mr}.

\subsection{The DECADE + DES-Y3 weak lensing catalog}
\label{sec:decade}

\begin{figure}
\includegraphics[width=\linewidth]{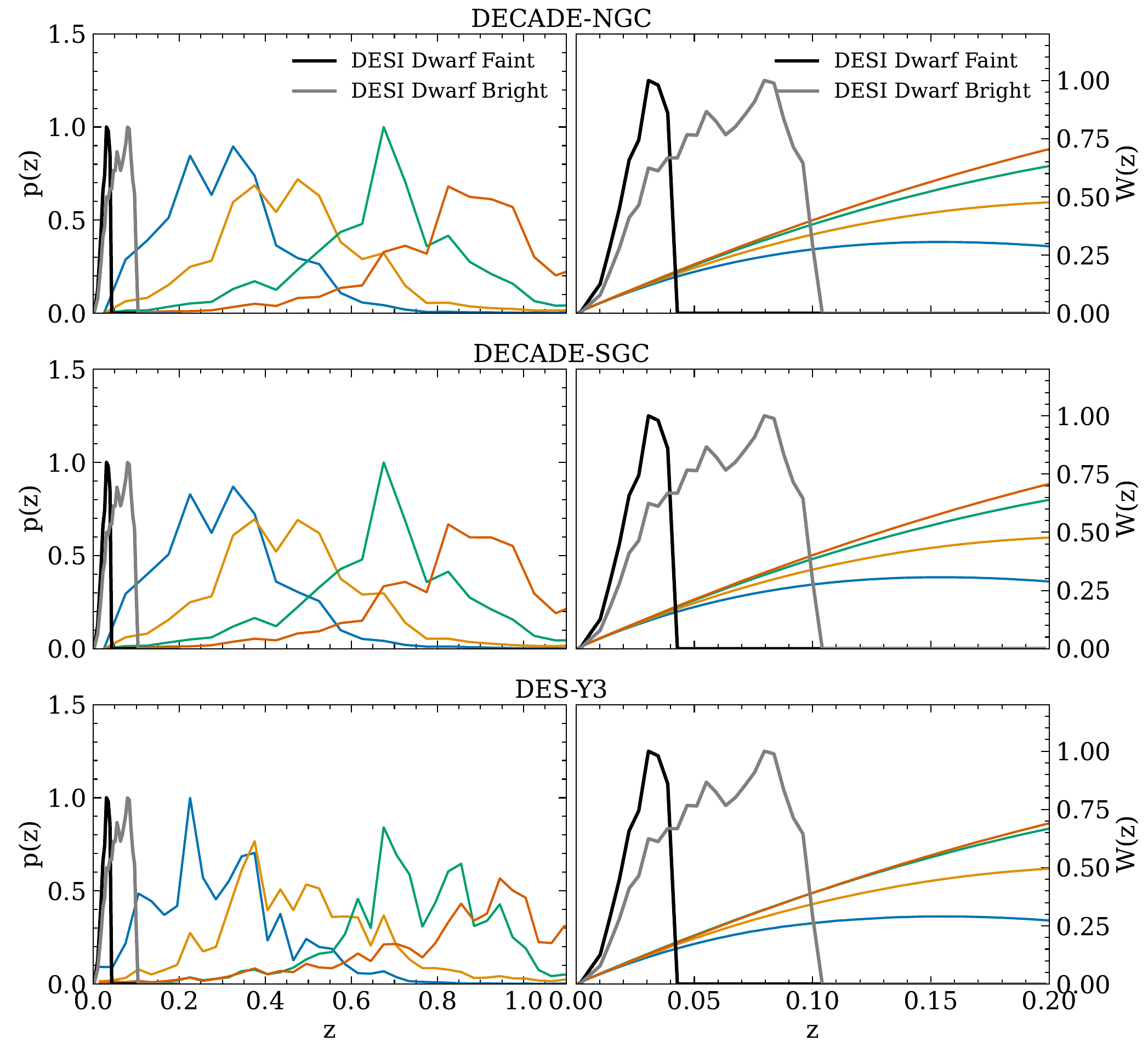}
\caption{Redshift distributions (left) and lensing efficiency (right) of the source galaxy samples. Different colors correspond to different tomographic bins. In both columns, we overplot the redshift distribution of the lens galaxy sample in black and grey. The product of the redshift distribution and the lensing efficiency is proportional to the lensing signal. Each curve is normalized so that the largest value across all redshifts is $1$. Three rows show three lensing datasets used in this analysis: DECADE-NGC (top), DECADE-SGC (middle), and DES-Y3 (bottom).}
\label{fig:nz}
\end{figure}

We use the combined weak lensing shear catalogs from the DES-Y3 \citep{y3-shapecatalog} and DECADE \citep{Anbajagane2025e}. The DECADE dataset further splits into the North Galactic Cap (NGC) and South Galactic Cap (SGC) regions, where the data from the NGC region was taken first --- these two regions were processed and calibrated\footnote{Here, we mean the calibration of shape measurement and redshift distribution.} separately. This combined sample covers a total of $\sim$13,000 degrees$^2$ and $\sim$270 million galaxies, and was also used in \citep{Anbajagane2025e} and \citep{marcomassmap}. Both DES and DECADE are based on imaging data from the Dark Energy Camera \citep[DECam,][]{Flaugher2015} and processed through a similar imaging pipeline \citep{Morganson2018,Tan2025, Anbajagane2025a}. Galaxy shapes in both sets of data were estimated and calibrated using the \texttt{Metacalibration} algorithm \citep{Sheldon2017,HuffMcal2017}. The final uncertainties in the shear calibration were estimated through image simulations in \citep{y3-imagesims} and \citep{Anbajagane2025a}, respectively. The characteristics of each of the catalogs are listed in Table~\ref{tab:numgal}. 

Galaxies in DES-Y3, DECADE-NGC, and DECADE-SGC are divided into four tomographic bins and calibrated using a self-organizing maps method as described in \citep{y3-sompz} and \citep{Anbajagane2025b}. We show the redshift distribution for each sample in Figure~\ref{fig:nz}. When performing our dwarf lensing measurements, we construct the estimator for each tomographic bin and dataset separately, and combine them into one data vector in the end (see Section~\ref{sec:measurement}). In Fig.~\ref{fig:nz}, the similarity of the redshift distribution between DECADE-NGC and DECADE-SGC is because they both rely on the same redshift training samples. The DES-Y3 analysis uses a slightly different redshift training set and adopts a finer redshift binning when producing the redshift distribution, resulting in different redshift distributions compared to DECADE.

\begin{table*}

\caption{Characteristics of source galaxies in four redshift bins. $n_{\rm{eff}}$ is the effective number density per square arcmin and $\sigma_e$ is the shape noise. Both are calculated using the definition of \citep{H12}. }
\begin{tabular}{ l l cc cc cc cc}
\hline
 Dataset & Fractional  & \multicolumn{2}{l}{Bin 1}  & \multicolumn{2}{l}{Bin 2} & \multicolumn{2}{l}{Bin 3} & \multicolumn{2}{l}{Bin 4} \\
  & overlap & $\sigma_{e}$ & $n_{\rm eff}$ & $\sigma_{e}$ & $n_{\rm eff}$ &  $\sigma_{e}$ & $n_{\rm eff}$  & $\sigma_{e}$ & $n_{\rm eff}$  \\ 
 \hline
 DES-Y3 &0.08 & 0.243 & 1.476 & 0.262 & 1.479  & 0.259 & 1.484 & 0.301 & 1.461 \\
 DECADE-NGC &0.72 & 0.233 & 1.239  & 0.259 & 1.150 & 0.248 & 1.169  &0.289 & 1.153 \\
 DEACDE-SGC &0.20  & 0.234 & 1.174 & 0.262 & 1.084 & 0.251 & 1.102 & 0.292 & 1.089 \\
\hline

\end{tabular}
\label{tab:numgal}
\end{table*}

\section{Measurement and modeling}
\label{sec:measurementandmodeling}

\subsection{The excess surface mass  density estimator $\Delta \Sigma$}
\label{sec:measurement}

The excess surface mass density ($\Delta \Sigma$) is measured using \textsc{dsigma}(v1.1.0) code \citep{2022ascl.soft04006L}\footnote{https://github.com/johannesulf/dsigma. We note that \textsc{dsigma} v1.1.0 had a problem computing the selection response for the DES sample that we fixed in this paper. Specifically, the tomographic bin assignment for each \textsc{metacal}-sheared sample is different and should be used when computing the selection response.}. The estimator used in this paper has been covered in several previous works \citep[e.g.][]{2024PhRvD.110h3509B, 2025arXiv250621677H}. Here we summarize the key steps and highlight the main differences. 

Our estimator for $\Delta \Sigma$ consists of three main components. The first is the weight assigned to each lens--source pair, which depends on the inverse critical surface density,
\begin{eqnarray}
\Sigma^{-1}_{\rm{crit}} (z_l, z_s) = \frac{4\pi G}{c^2} \frac{D(0, z_l)}{D(0, z_s)} \max(0, D(z_l, z_s)),
\end{eqnarray}
where $c$ is the speed of light, $G$ is the gravitational constant, and $D(x, y)$ is the angular diameter distance between redshifts $x$ and $y$. However, this formulation requires knowledge of the redshift for each source galaxy, which is typically not available with sufficient precision. To address this, we instead use the average lensing efficiency for the tomographic bin to which the source belongs:
\begin{equation}
\langle \Sigma^{-1}_{\rm{crit}} \rangle_{l,s} = \int \Sigma^{-1}_{\rm{crit}} (z_l, z_s)  n(z_s) , dz_s,
\end{equation}
where $n(z_s)$ is the redshift distribution of the tomographic bin. 

Second, the lensing shear is measured using the \textsc{Metacalibration} technique. Specifically, the mean tangential shear around a lens $l$, denoted $\langle \bm{\gamma_t} \rangle_l$, is related to the mean tangential ellipticity of source galaxies ($\langle \bm{e_t} \rangle_l$) via:
\begin{eqnarray}
\langle \bm{\gamma_t} \rangle_l &=& \left(\langle R_t \rangle_l + \langle R_s \rangle\right)^{-1} \langle \bm{e_t} \rangle_l, \\
\langle R_t \rangle_l &=& \langle R_{11} \cos^2(2\phi) + R_{22} \sin^2(2\phi) \nonumber \\
&&+ (R_{12} + R_{21}) \sin(2\phi)\cos(2\phi) \rangle,
\end{eqnarray}
where $R_{ij}$ are components of the shear response matrix, $\phi$ is the polar angle of the source galaxy relative to the lens, and $\langle...\rangle$ denotes an average over source galaxies. The term $\langle R_s \rangle$ is the selection response from \textsc{Metacalibration}, computed separately for each tomographic bin.

Finally, some source galaxies reside at the same redshift as the lens galaxies and thus do not experience gravitational shear. Their inclusion dilutes the lensing signal. To correct for this bias, we apply a multiplicative factor known as the boost factor, $B(r)$, which accounts for the excess number of galaxies physically associated with the lens. Typically, the boost factor is estimated via the ratio of source--lens galaxy pairs to the source--random position pairs \citep{sheldon2004}. This method requires a random catalog for the lens galaxy and has been shown in simulations that it can be biased \citep{2019MNRAS.489.2511V}.

Instead, we adopt the ``$P(z)$ decomposition'' method  \citep{2019MNRAS.489.2511V,2021PhDT........16P} to determine the boost factor, which requires redshift estimates for individual source galaxies. This method has been widely used in cluster lensing studies \citep{2019MNRAS.489.2511V,2021PhDT........16P,SPT2023,2024A&A...687A.178G}; here we briefly summarize the key concepts. We first obtain photometric redshifts for each source galaxy using the Directional Neighborhood Fitting (DNF) algorithm \citep{DNF}, similar to \citet{SPT2023}. For each lens, we compute the redshift distribution of source galaxies in annuli of radius $r$, denoted $P(z_s|r)$. We model this distribution as a mixture of two components: one associated with galaxies at the same redshift as the lens, and one representing background galaxies, modeled using the distribution at large radius ($r_{10}=10h^{-1}\mathrm{Mpc}$):
\begin{eqnarray}
P(z_s|r) &=& f_l(r) \mathcal{N}(\mu, \sigma) \nonumber \\
&&+\left(1 - f_l(r)\right) P(z_s|r_{10}),
\end{eqnarray}
where $\mathcal{N}(\mu, \sigma)$ is a Gaussian distribution with mean $\mu$ and scatter $\sigma$, representing physically associated sources, 
and $f_l(r)$ is the fraction of sources in the same redshift as the lens. The free parameters in the model include $\mu, \sigma$, and the value of $f_l(r)$ at each radius bin. The boost factor is then given by:
\begin{equation}
\label{eq:boost}
B(r) = \frac{1}{1 - f_l(r)}.
\end{equation}
Note that the $P(z)$ decomposition only relies on the relative redshifts between source galaxies in different line of sight.  
The requirement on the absolute precision of the redshift estimation is less stringent. 

Combining all three components, our final estimator for the excess surface mass density is:
\begin{eqnarray}
    \Delta \Sigma_l (r) &=&\frac{\sum_{ls} w_l w_s \langle \Sigma^{-1}_{\rm{crit}}\rangle_{l,s} \bm{e_{t,s}}}{\sum_{ls} w_l w_s \langle \Sigma^{-1}_{\rm{crit}}\rangle_{l,s}^2}\nonumber\\ 
    &&\times \frac{\sum_{ls} w_l w_s \langle \Sigma^{-1}_{\rm{crit}}\rangle_{l,s}^2}{\sum_{ls} w_l w_s (\langle R_t \rangle_l + \langle R_s\rangle)\langle \Sigma^{-1}_{\rm{crit}} \rangle_{l,s}^2}\nonumber\\
    &&\times \frac{1}{B(r)}, 
\end{eqnarray}
where $w_s$ are  the source weights provided by the shape catalogs, $w_l$ is $w_{\rm{zfail}} w_{\rm{IIP}}$, and $\bm{e_t}$ is the tengential shapes of source $s$ around lens $l$. In the above equation, the first term is the minimal variance estimator of $\Delta \Sigma$. The second term is the weighted mean of the \textsc{Metacalibration} shear response. The third term is the boost factor correction derived using the $P(z)$ decomposition method. 

\begin{figure*}
\includegraphics[width=\linewidth]{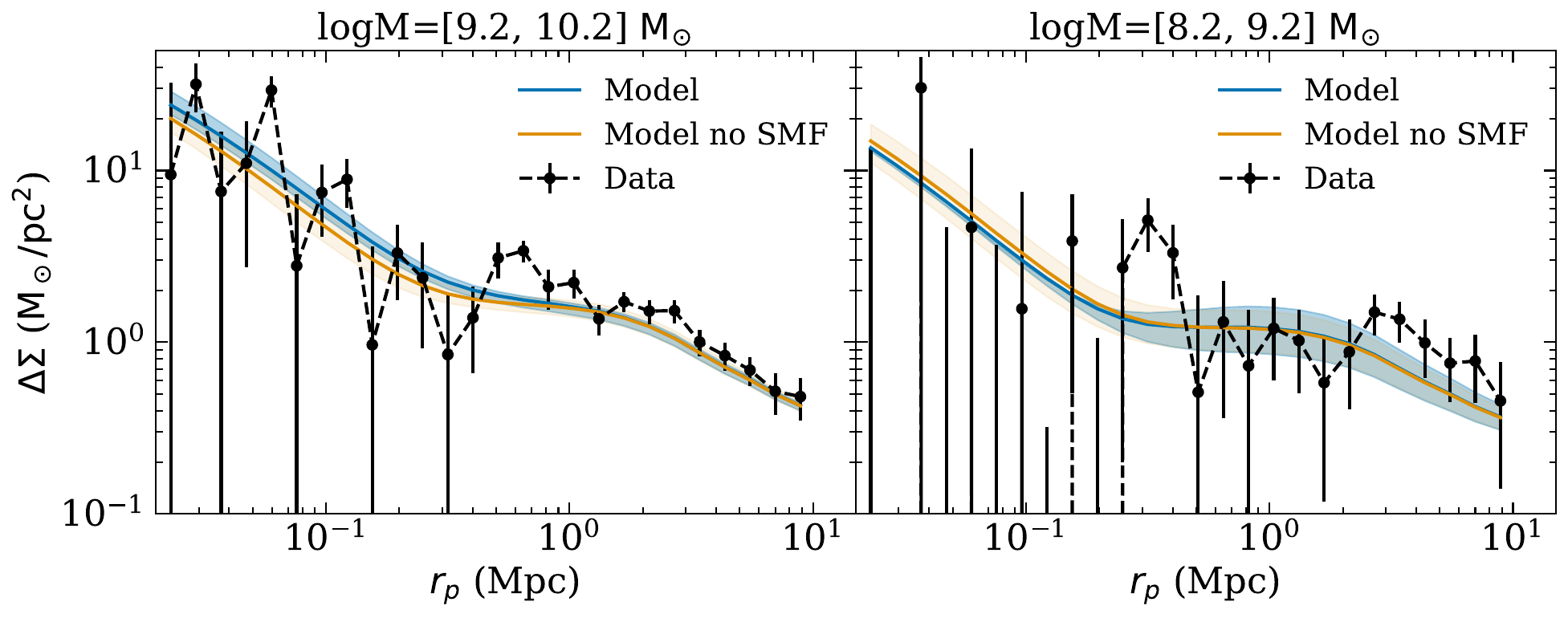}
\caption{Comparison of the model and the measured weak lensing profiles around dwarf galaxies in two stellar mass bins. Dots with the $1\sigma$ error bars show the measurements. Lines show the best-fit model with (blue) and without (brown) stellar mass function priors. Color bands indicate $68\%$ credible intervals.}
\label{fig:delta_sigma}
\end{figure*}

\subsection{Forward-modeling}
\label{sec:model}

We employ the gravity-only N-body simulation IllustrisTNG-300 Dark \citep{TNG} to model the measurement. IllustrisTNG-300 Dark has a volume of $205^3\ h^{-1}\rm{Mpc}^3$ with particle resolution $4.72 \times 10^7 h^{-1}\rm{M}_{\odot}$. The simulation is generated using \textsc{Arepo} \citep{2020ApJS..248...32W} with a $\Lambda$CDM cosmology that has $\Omega_{\rm{m}}=0.3089$,  $\Omega_{\rm{b}}=0.0486$, $h=0.6774$, $n_{\rm{s}}= 0.9667$, and $\sigma=0.8159$. We perform the halo finding using \textsc{Rockstar} (v0.99.9-RC3+) \citep{Rockstar} and generate merger trees using  \textsc{Consistent Trees}  (v1.01) \citep{consistentree}. In the main analysis, we use the snapshot at $z=0.07$, close to the mean redshift of the high stellar mass sample ($=0.073$). We validate the analysis with snapshot $z=0.03$, close to the mean redshift of the low stellar mass sample ($z=0.031$), and find negligible differences.

Following \citep{Thornton}, we connect the stellar mass ($M_*$) to the peak halo mass ($M_{\rm{peak}}$) using a simple log-normal model 
\begin{align}
\label{eq:stellarmasshalomass}
    &P(\log M_* | \log M_{\rm peak}) \notag \\
    &= \mathcal{N}\left(\log M_{11}+\alpha(\log\left(\frac{M_{\rm peak}}{h^{-1}\Msun}\right)-11), \sigma_{\log M}\right),
\end{align}
where $\log M_{11}$, $\alpha$, and $\sigma_{\log M}$ are free parameters.  Our model assumes that satellite and central galaxies follow the same $M_*$–$M_{\rm{peak}}$ relation. This assumption may be too restrictive; for example, \citep{reddick2013} demonstrates that different choices for the stellar mass–halo property relation can lead to significant variations in the predicted satellite fraction $f_{\rm{sat}}$. Moreover, the baryonic component of the main halo disrupts dwarf galaxies during pericentric passages, reducing the satellite fraction \citep{2018ApJ...859..129N}.

To account for these effects, we introduce two free parameters, $f_{\rm{sat}, h}$ and $f_{\rm{sat}, l}$, which describe the mean satellite fraction in the high- and low-stellar-mass bins, respectively. We identify satellites in the simulation using the \textsc{Rockstar} halo catalog, based on the unique parent ID (UPID). After compiling the list of satellites, we randomly downsample it to match the satellite fractions specified by $f_{\rm{sat}, h}$ and $f_{\rm{sat}, l}$ in the corresponding stellar mass bins.

Operationally, there is one subtlety in constructing the satellite list. In this scheme, subhalos with $\rm{UPID} \neq -1$ are normally classified as satellites. However, this definition depends on the percolation algorithm used by \textsc{Rockstar}, which only links subhalos within a specific phase-space distance, known as the percolation radius \citep{2019MNRAS.489.4170G}. As a result, heavily stripped subhalos, often known as Splashback subhalos \citep{More15,Susmita14,Diemier14}, that fall outside this percolation radius may be assigned $\rm{UPID} = -1$ despite being satellites. To mitigate this issue, we additionally classify any halo with $M_{\rm{peak}} / M_{\rm{now}} > 1.01$ as a satellite, even if it has $\rm{UPID} = -1$. In the fiducial analysis, we do not consider orphan subhalos. We demonstrate that the inclusion of orphan subhalos does not significantly change our results in section \ref{sec:results}.

Finally, to predict the $\Delta \Sigma$ profile, we first measure the projected mass density $\Sigma(r)$ using the dark matter particles for each subhalo and main halo in the catalog. We then transform $\Sigma(r)$ to $\Delta \Sigma (r)$ using 
\begin{eqnarray}
    \Delta \Sigma(r) = \langle \Sigma(<r) \rangle-\Sigma(r),  
\end{eqnarray}
where $\langle...\rangle$ indicates an average over angular scales. This process is performed using \textsc{halotools} \citep{halotool}. For each MCMC step, we select all halos within the stellar mass bin. We then randomly downsample the subhalos so that the satellite fraction matches the parameters. Finally, we measure the average of $\Delta \Sigma$ for the selected halos and subhalos. We repeat the above processes 20 times and take the average of the $\Delta \Sigma$ to decrease the stochasticity at each MCMC step.

\begin{table*}
\caption{Summaries of the marginalized parameter constraints from simulation-based models. Values show the mean of the marginalized posterior and the $68\%$ credible intervals estimated by the \textsc{GetDist} package \citep{getdist}.}
\centering
\begin{tabular} { l  c c}
\noalign{\vskip 3pt}\hline\noalign{\vskip 1.5pt}\hline\noalign{\vskip 5pt}
 \multicolumn{1}{c}{\bf } &  \multicolumn{1}{c}{\bf Constrained value} &  \multicolumn{1}{c}{\bf Descriptions}\\
\noalign{\vskip 3pt}\cline{2-3}\noalign{\vskip 3pt}
\hline
{\boldmath$\log M_{11}    $} & $8.53^{+0.51}_{-0.18}      $ & mean log stellar mass in unit of log $M_\odot$ when halo mass is $10^{11} h^{-1} \Msun$  \\

{\boldmath$\alpha         $} & $1.79^{+0.58}_{-0.19}      $ &  slope of stellar mass--halo mass relation\\

{\boldmath$\sigma_{\log M}$} & $0.71^{+0.49}_{-0.36}      $ &  scatter of stellar mass--halo mass relation\\

{\boldmath$f_{\rm{sat}, h}$} & $0.343^{+0.070}_{-0.097}   $ & satellite fraction of galaxies with $M_*=[10^{9.2}, 10^{10.2}] \Msun$\\

{\boldmath$f_{\rm{sat}, l}$} & $0.283^{+0.097}_{-0.16}    $ & satellite fraction of galaxies with $M_*=[10^{8.2}, 10^{9.2}] \Msun$\\
\hline
\end{tabular}

\label{tab:param}
\end{table*}
\begin{figure*}
\includegraphics[width=0.8\linewidth]{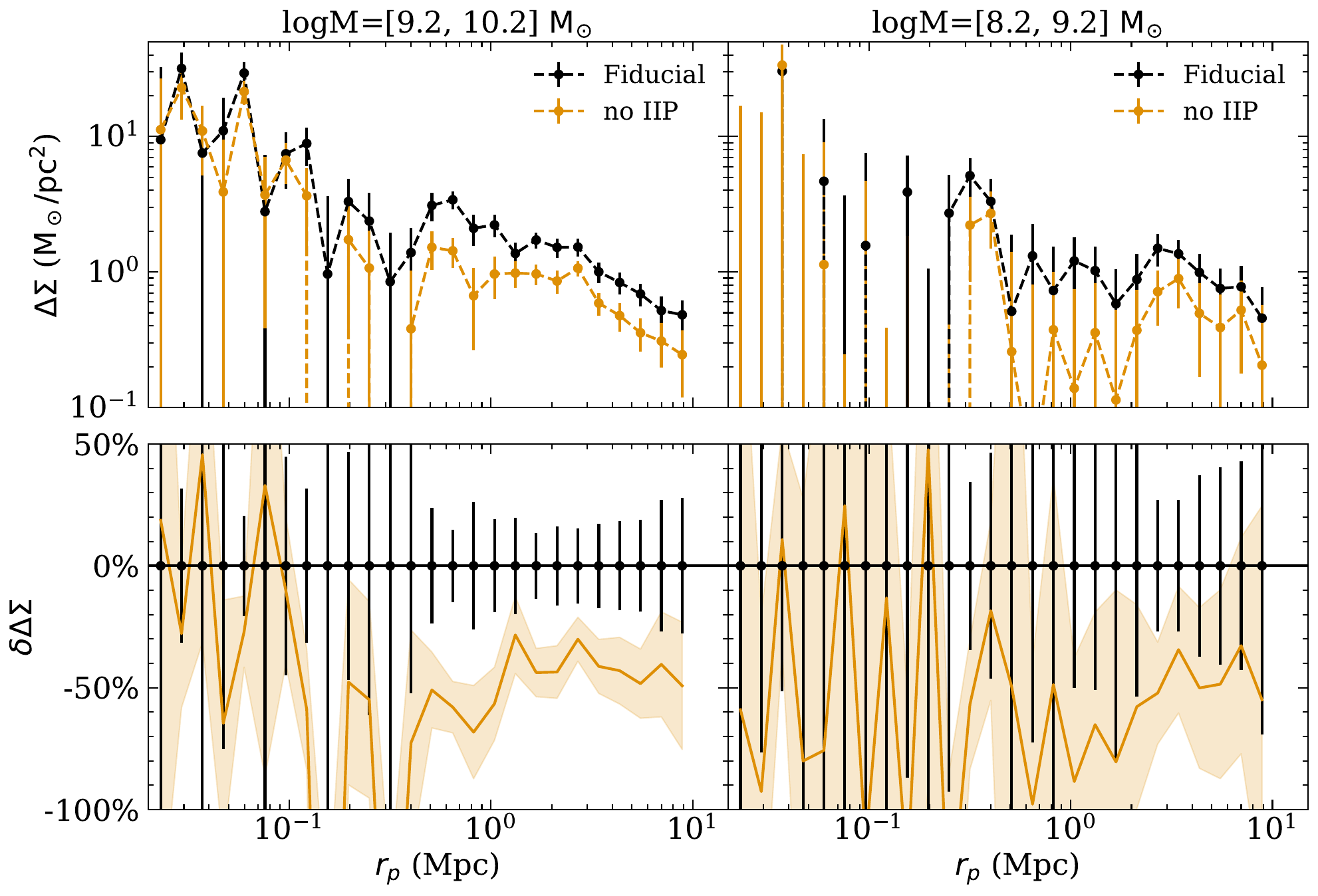}
\caption{Comparison of the lensing measurements of galaxies in two stellar mass bins with and without IIP weights. Error bars show $1\sigma$ uncertainties. Black dots correspond to the fiducial analysis, while the brown dots show alternative stellar mass estimates. The bottom panels show the fractional differences to the fiducial measurements. The error bars and shaded regions show the $1\sigma$ uncertainties.}
\label{fig:weight}
\end{figure*}
\subsection{Data vector and Likelihood inference}
We adopt a Gaussian likelihood framework, where the model predictions ($M$) are described in Section~\ref{sec:model} and the measurements ($d$) are detailed in Section~\ref{sec:measurement}. The covariance matrix $C_{\rm{jk}}$ is estimated using the jackknife resampling method with 100 patches, each covering approximately $43\ \mathrm{deg}^2$. This patch size is about three times larger than the maximum angular scale used in our analysis. The finite number of jackknife patches introduces noise in the estimated covariance matrix, which in turn biases its inverse. To correct for this, we apply the Hartlap correction \citep{2007A&A...464..399H}, modifying the inverse covariance matrix as follows,
\begin{eqnarray}
C^{-1} = \frac{n - p - 2}{n - 1} C_{\rm{JK}}^{-1},
\end{eqnarray}
where $n$ is the number of Jackknife patches and $p$ is the dimensionality of the weak lensing data vector $d$.

We measure $\Delta \Sigma$ in $30$ equally log-spaced comoving radial bins from $0.01$ to $10\ \rm{Mpc}$. We cut the scales below $30\ h^{-1} \rm{kpc}$, which is roughly $15$ times the softening length of the simulation.

Due to the relatively low signal-to-noise ratio of the galaxy--galaxy lensing measurements, we observe a strong degeneracy between the parameters $\alpha$, $\log M_{11}$, and $\sigma_{\log M}$. To break this degeneracy, we introduce an external prior based on the stellar mass function. Specifically, we compute the expected number density of galaxies ($n_d$) in each stellar mass bin using the stellar mass function measured in DESI DR1 \citep{2025MNRAS.540.1635X}, and compare it to the number density predicted by the simulation ($n_{\rm{sim}}$) at each step of the MCMC.

To conservatively account for potential systematics in the stellar mass function, we adopt a $20\%$ uncertainty on $n_d$, which is roughly 20 times larger than the reported statistical uncertainty. This $20\%$ encompases some differences of stellar mass function estimations in the literature. For example, \citep{2024ApJ...971..119W} measures the stellar mass function with Legacy Survey DR9, DESI SV3, and DESI Y1 data, finding a value different from \citep{2025MNRAS.540.1635X} at $8\%$ levels. While we think this prior is relatively conservative, we note that our result does not hinge on the validity of this prior. For all the main results below, we show the fits without this prior and find consistency.  
The final log-posterior is given by:
\begin{eqnarray}
\log P(d|M) &=& -\frac{1}{2} (d - M) C^{-1} (d - M)^T  \notag \\
&& + \log U(n_d,\ 0.8n_{\rm{sim}},\ 1.2n_{\rm{sim}}),
\end{eqnarray}
where the uniform prior $U(x, y, z) = 1$ if $x \in [y, z]$, and $0$ otherwise. This posterior is sampled using the affine-invariant ensemble MCMC sampler implemented in \textsc{emcee} \cite{emcee}. Full sampled posteriors are shown in Fig.~\ref{fig:contour}.

\section{Results}
\label{sec:results}

Figure \ref{fig:delta_sigma} shows a comparison between the measurement and the model. We find a signal-to-noise of $12.4$ and $5.6$ for the $\Delta \Sigma$ measurements of two stellar mass bins, respectively, whose mean stellar mass is $10^{9.44\pm0.18}\ \Msolar$ and $10^{8.65\pm 0.28} \Msolar$, respectively. To the best of our knowledge, this is the first detection of lensing around spectroscopically confirmed dwarf galaxy samples. We further discuss the nature of this measurement in section \ref{sec:detection}. Our model fits the measurement with a probability-to-exceed (PTE) value of $0.03$. We note that this PTE value drops to $0.008$ when two $f_{\rm{sat}}$ parameters are not included in the model, indicating the need to include those two parameters. We further compare the model fit without imposing a stellar mass function prior in Fig.~\ref {fig:delta_sigma} and find consistency with our fiducial result. Specifically, the change of the best-fit $\chi^2$ is $\simeq 2$ with degree-of-freedom of $47$. The sampled posteriors are shown in Fig.~\ref{fig:contour} and the marginalized constraints are shown in Table~\ref {tab:param}. Below, we discuss the robustness of the measurements and their scientific implications.

\subsection{Measurement robustness}
\label{sec:mr}

We test whether our measurements are robust to three factors in the analysis: the stellar mass estimate, the shredding of faint objects, and the fiber incompleteness.

As described in section \ref{sec:stellarmass}, stellar mass estimation of the dwarf galaxies is challenging. Different assumptions in the SED fittings can lead to a shift of at least $0.15$ dex. We test the robustness of the measurement using three stellar mass estimates: one photometry-based (SAGA) and two SED-based methods (CIGALE and CIGALE-noAGN). To ensure an apples-to-apples comparison, we compare the measurement in an abundance-matched way. Specifically, we make a stellar mass cut using the CIGALE-noAGN mass. We then define the stellar mass cut for CIGALE and SAGA stellar mass so that the number of dwarf galaxies above the stellar mass cut is the same as that of the cut based on CIGALE mass. We then remeasure the $\Delta \Sigma$ profile for the newly selected samples. We find a $\Delta \chi^2\simeq 10$ with degree-of-freedom of $52$ between $\Delta \Sigma$ generated with SAGA stellar mass and CIGALE-noAGN stellar mass. The $\Delta \chi^2 $ goes up to $\simeq 34$ when comparing CIGALE-noAGN and CIGALE. We investigate further the differences between CIGALE-noAGN and CIGALE in Appendix \ref{sec:stellar_mass} and conclude that CIGALE's AGN fraction might be overestimated, which might explain the differences in stellar mass estimates. We leave detailed investigations to future work. 

Dwarf galaxies are often spatially extended, making them likely to be shredded into multiple objects \citep{Elise, 2023ApJS..269....3M,2012AJ....143..133H}. This makes accurate photometric measurement particularly challenging. To address this, we limit the main analysis to \texttt{FRACFLUX}$<0.2$ in $g,r,z$ bands, which quantifies the fraction of the flux from nearby objects. We find negligible impacts of this cut to the result presented in this paper (Figure~\ref{fig:fracvut}). 

Finally, we test the impact of IIP weights on the measurement.  Figure \ref{fig:weight} shows the comparison between $\Delta \Sigma$ with and without IIP weights. Consistent with \citep{desilensing}, we find that the $\Delta \Sigma$ measurements without IIP weights are preferentially lower than those with IIP weights. Since the fiber assignment preferentially downweights overdense regions, the halo mass and the satellite fraction in the samples will be preferentially lower. The lower satellite fraction will lead to a lower $\Delta \Sigma$ on large scales (see fig. 10 of \citep{Thornton}), consistent with our observations. The lower halo mass leads to a smaller one-halo term, which is less obvious in Fig.~\ref{fig:weight}. However, when fitting to the measurement without IIP weights, we observe a shift in the stellar mass--halo mass relation (Figs.~\ref{fig:contour} and \ref{fig:smhm_noiip}). As we will discuss below, \citep{desilensing} demonstrates in mock catalogs that the IIP weights calibrate the incompleteness to the $1\%$ level in the lensing profiles, and our final results are insensitive to calibration errors that are thrice as large. We, therefore, adopt the measurement with IIP weights as our fiducial analysis without inflating error bars corresponding to IIP weight uncertainties. Finally, we note that the redshift failure weight ($w_{\rm{zfail}}$) has a negligible impact on the measurement. 

In Appendix \ref{sec:alllensingsys}, we test other lensing systematics, including multiplicative biases, boost factor corrections, and lensing B-mode, and find that these systematics do not change the conclusion of this paper.  

\begin{figure}
\includegraphics[width=\linewidth]{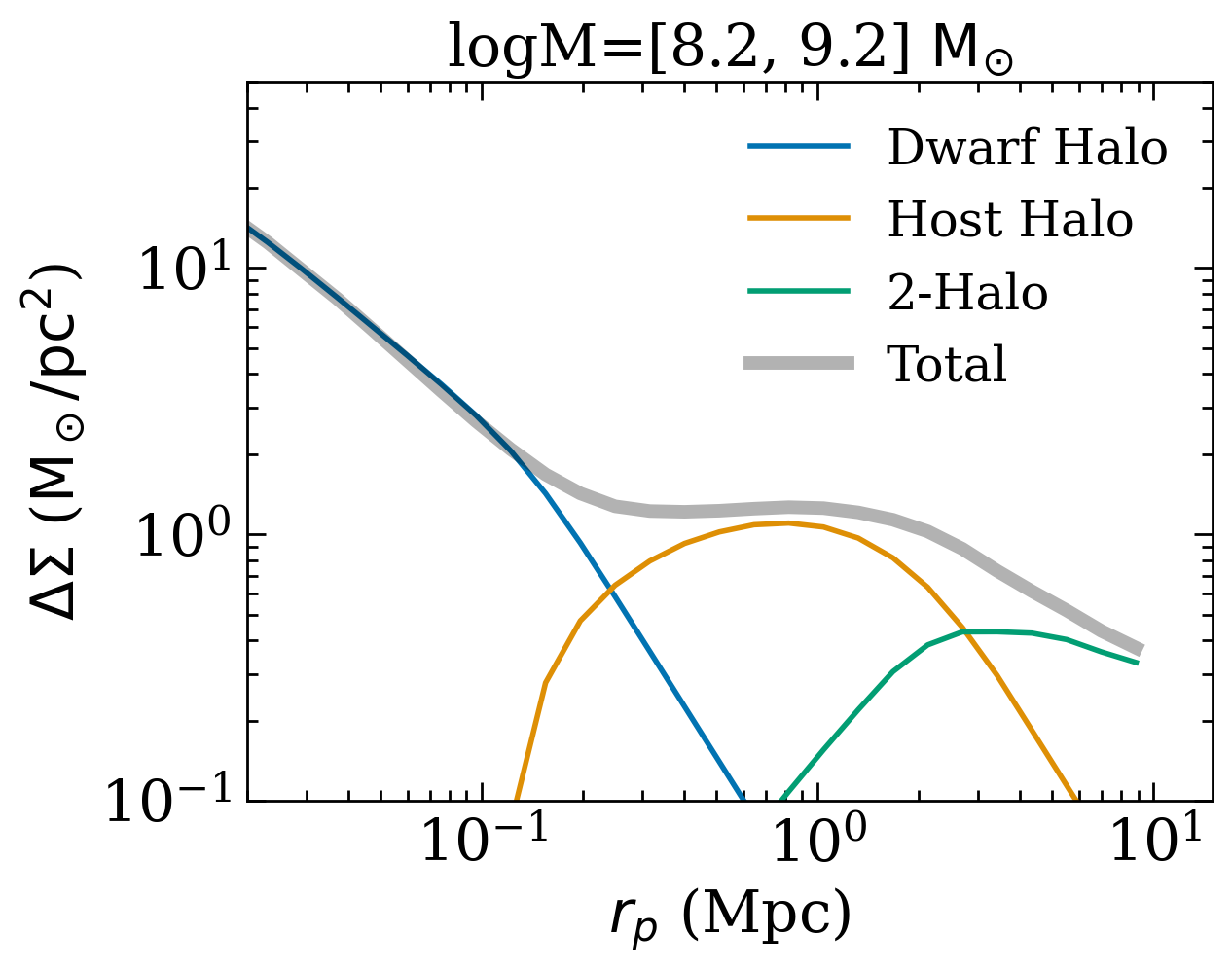}
\caption{A breakdown of the best-fit model for the low stellar mass bin. The black line shows the total lensing profile. The blue line shows the contribution from the host halos of the dwarf galaxies. The brown line shows the contribution from the main halos in which the dwarf galaxies reside as satellites. In this calculation, we exclude the matter in subhalos that contributes to the blue line to avoid double-counting. The green line shows the matter outside the main halos, known as the two-halo term. In the above calculations, we use the virial radius of the halo as the halo boundary.}
\label{fig:breakdown}
\end{figure}
\subsubsection{The nature of the detection}
\label{sec:detection}
Since the fiber incompleteness alters the galaxy selection function that is correlated with the environment, the signal-to-noise changes when we apply the IIP weight to mitigate this issue. Without applying the IIP weight, the signal-to-noise of the measurement drops to $9.2$ and $4.5$ for high and low stellar mass bins, respectively. 

The measured lensing profile around galaxies encompasses three contributions: (1) the matter in the halos that host the dwarf galaxies (dwarf halo hereafter, the blue line in Fig.~\ref{fig:breakdown}), (2) the matter in the main halos of which the dwarf galaxies are satellites (host halo hereafter, the brown line in Fig.~\ref{fig:breakdown}), and (3) the matter outside of the main halos, known as the two-halo term (the green line in Fig.~\ref{fig:breakdown}). For the discussion below, we focus on the low mass bin, which corresponds to dwarf galaxies. First, to remove the contribution from the two-halo term, we remove the scale that corresponds to the Splashback radius of the halos, calculated using the formula in \citep{2020ApJS..251...17D} and the mean halo mass estimated from the lensing profile. The signal-to-noise drop from $5.6$ to $3.8$ after we remove the two-halo contribution. Further, we want to understand whether the remaining signal-to-noise comes from the matter in the halos that host dwarf galaxies or from the host halos. It is challenging to disentangle the contributions from these two sources. However, conveniently, the measurement without IIP weights has a low satellite fraction due to fiber incompleteness. Therefore, the measurement without IIP weights should have a low contribution from the host halos. The signal-to-noise of this measurement after removing two-halo contributions can serve as an approximation of the signal-to-noise from the dwarf halo. We find the signal-to-noise of this measurement $3.5$.

\begin{figure*}
\includegraphics[width=0.8\linewidth]{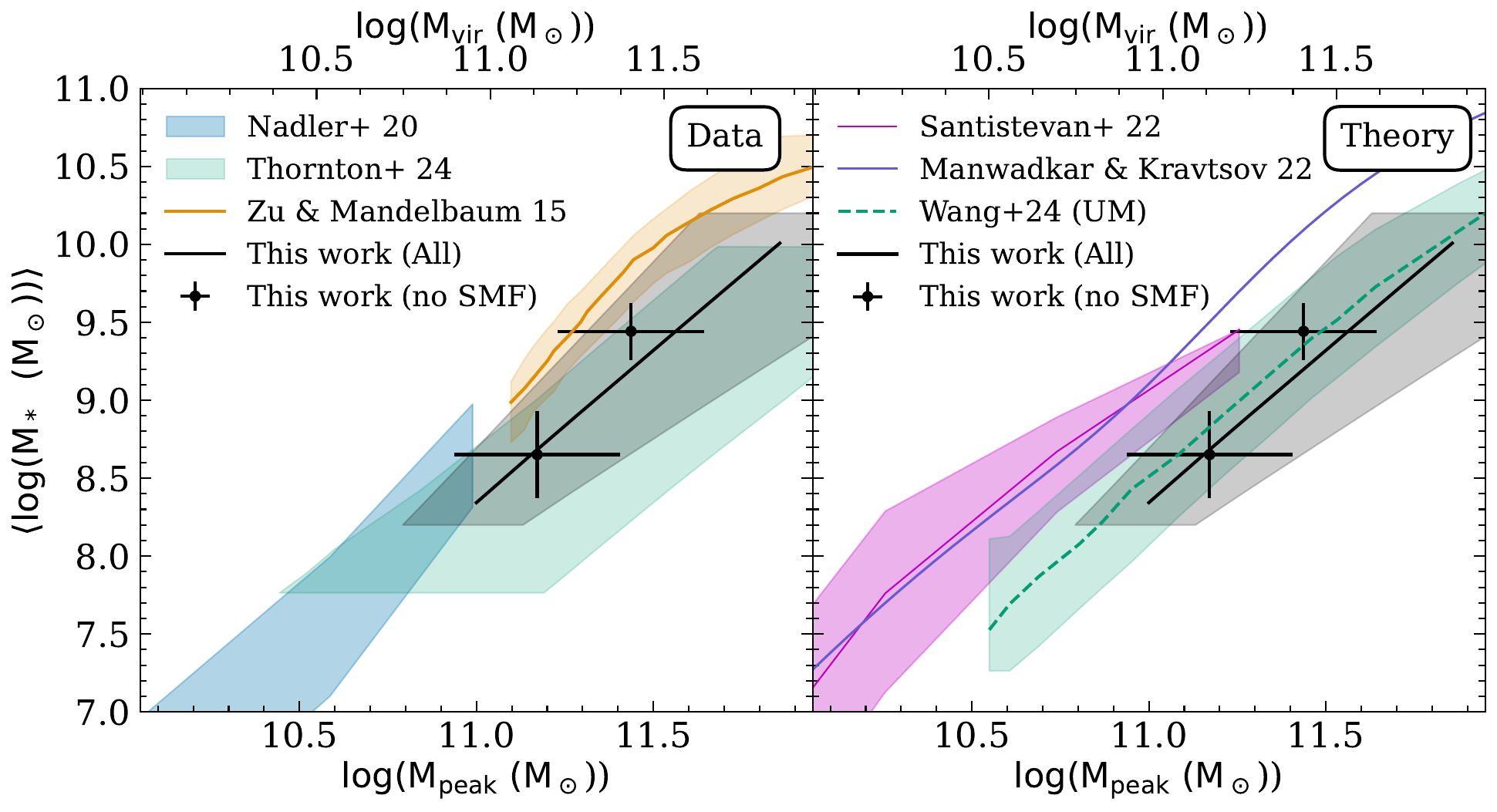}
\caption{The stellar mass--halo mass relations constrained from the model and their comparisons with the measurements in the literature (left) and theory predictions (right). The black line shows the best-fit value of the fiducial analyses, and the shaded black regions show the $68\%$ credible intervals. For more model-independent constraints, we show black dots with error bars. The x-axis of the dots shows the best-fit halo mass from the analysis of lensing profiles without applying the priors derived from stellar mass functions. The error bars show the corresponding $68\%$ credible intervals. The y-axis of the black dots corresponds to the mean and standard deviation of the stellar mass of dwarf galaxies in the two stellar mass bins considered in this analysis. We overlay the constraints and theory predictions from different literatures. Left: the brown line shows constraints from the joint analysis of galaxy--galaxy lensing and galaxy clustering in SDSS \citep{ZuandMandelbaum} with shaded regions showing $1\sigma$ uncertainties. The green bands show $1\sigma$ constraints from the galaxy--galaxy lensing measurements around photometrically selected dwarf galaxies \citep{Thornton}. The blue shaded region shows constraints from Milky Way satellite abundances \citep{Nadler21}. Right: the pink lines with $1\sigma$ bands show predictions based on FIRE hydrodynamic simulations \citep{Fire}. The blue lines \citep{2022MNRAS.516.3944M} show predictions from a simple regulator-type model \citep{Grumpy} that fits to the constraints from \citep{Nadler21}. The green lines correspond to a semi-analytic model fitted to the SAGA satellites \citep{saga5}.
}
\label{fig:smhm}
\end{figure*}

\subsection{Stellar mass--halo mass relation}
\label{sec:smhm}
Figure \ref{fig:smhm} shows the inferred stellar mass--halo mass relation constraints from our posterior. The black lines and grey bands show the joint constraints from each stellar mass bin and the stellar mass function prior. The black dots with error bars show the individual fit without the stellar mass function prior, which is a more model-independent measurement of the stellar mass--halo mass relation. We find excellent agreement between the two. 

In the left panel of Figure~\ref{fig:smhm}, we compare our constraints to measurements in other literature. We find excellent agreement with \citet{Thornton}, which measured the galaxy--galaxy lensing signal around photometrically selected dwarf galaxies. Our constraints are also consistent with those from \citet{Nadler21}, which relies on the counts of Milky Way satellites. To connect these measured counts to halo mass, one needs to assume a model of the observable--halo mass relation and a simulation that can simulate the expected subhalo mass distribution that corresponds to the Milky Way system. In comparison, our results rely on galaxy--galaxy lensing, which measures the excess surface matter density profile around the objects. Connecting the measured excess surface matter density profile to the halo mass requires assumptions about the connection between the three-dimensional matter distribution and the projected matter distribution of the halo; arguably, it is less model-dependent. Compared to \citet{Thornton}, our result is based on spectroscopically selected dwarf galaxies, which are less affected by projections of background galaxies. Our results complement those measurements, and the consistency between our results and \citet{Thornton,Nadler21} is therefore an exciting confirmation. We also compare our result with \citet{ZuandMandelbaum}, which relies on galaxy clustering and galaxy--galaxy lensing around spectroscopically-selected galaxies in SDSS. Their lowest stellar mass sample overlap with the high stellar mass bin in this analysis. To make an apples-to-apples comparison, we measure the relation between $M_{\rm{peak}}$ and the halo viral mass $M_{\rm{vir}}$ for main halos in the IllustrisTNG-300 Dark simulation, and use it to translate our measurement of $M_*$--$M_{\rm{peak}}$ relation into $M_*$--$M_{\rm{vir}}$ relation. While our result is statistically consistent with their measurements, we find $\sim1\sigma$ shifts. We leave further investigations of this shift to future work. 

In the right panel of Figure~\ref{fig:smhm}, we compare our measurements to theoretical models and simulations. First, we find excellent agreement with \citet{saga5}, a tailored \textsc{UniverseMachine} galaxy--halo connection model \citep{universemachine} applied to cosmological zoom-in simulations and tuned to match the observed properties of dwarf galaxies in the SAGA survey \citep{saga3}. When comparing to models and simulations designed for Milky Way satellites (\citet{Fire, Grumpy}), we find a $\sim 1\sigma$ discrepancy. While the discrepancy is not statistically significant, we note that the majority of galaxies in our samples are not satellite galaxies. The stellar mass--halo mass relation might be different between satellites and isolated halos. We leave further investigations to future work.   %

\subsection{Satellite fraction}
\begin{figure*}
\includegraphics[width=\linewidth]{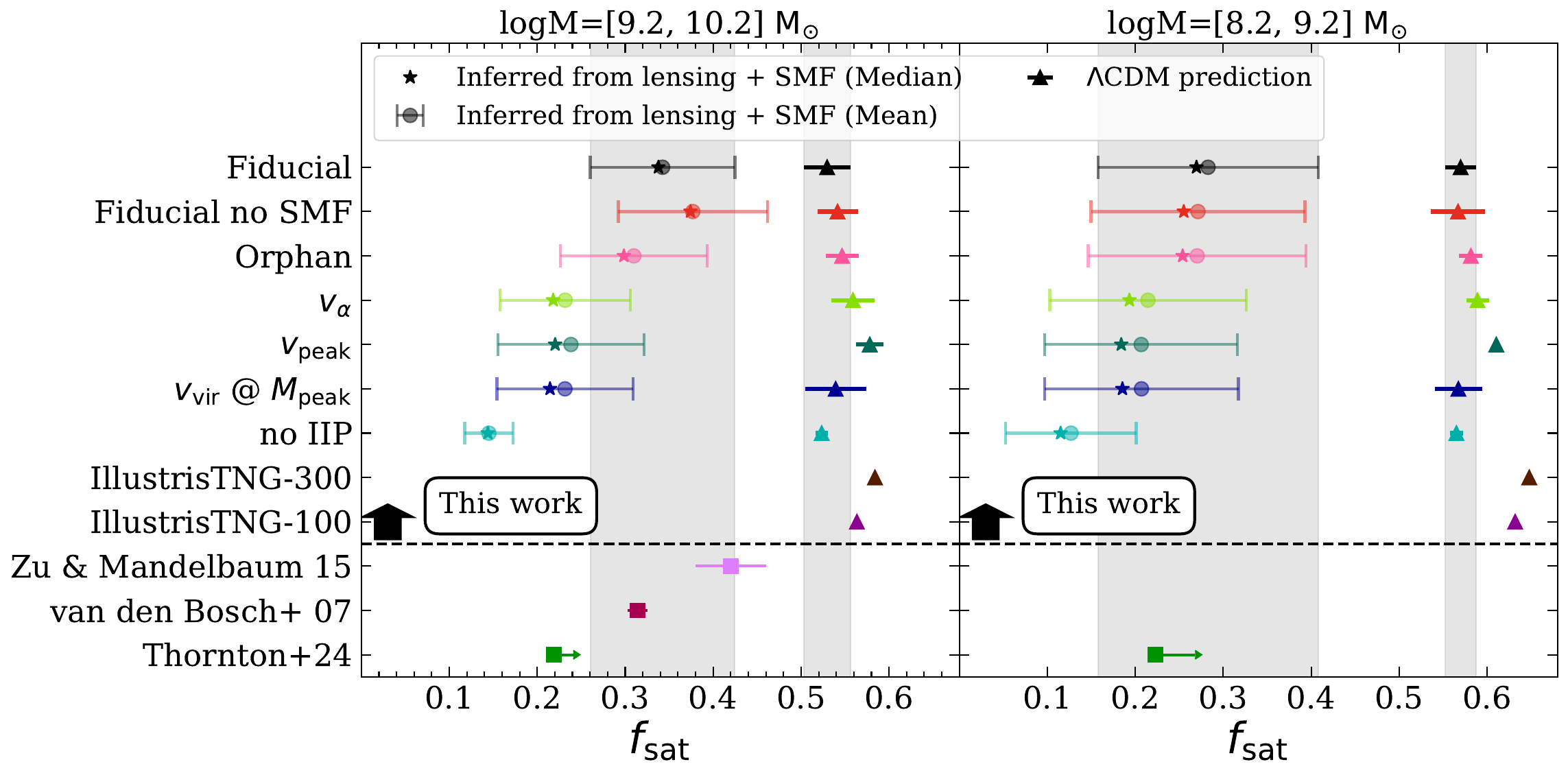}
\caption{Satellite fraction constraints for different analysis variants. Error bars indicate the $68\%$ credible intervals. Dots show the marginalized means from weak lensing profile analyses with stellar mass function priors, while stars show the median of the marginalized constraints. The triangles show the mean $\Lambda$CDM prediction based on the stellar mass–halo mass relation constrained from those same analyses. Each row corresponds to a different analysis variant, including removing the stellar mass function prior (red), considering orphan subhalos (pink), changing halo mass proxies (light green, dark green, and blue), and analyzing data with IIP weights (emerald). We further show predictions from the Illustris-TNG hydrodynamic simulation \citep{TNG} (seal brown and purple). We compare the constraints to those in the literature, including constraints from galaxy--galaxy lensing and galaxy clustering \citep{ZuandMandelbaum} (magenta), from galaxy group catalogs \citep{group08}, and from galaxy--galaxy lensing around photometrically selected dwarfs \citep{Thornton} (islamic green). Note that constraints from \citep{Thornton} are lower limits because of the potential incompleteness of sample selections.}
\label{fig:fsat}
\end{figure*}

Figure~\ref{fig:fsat} shows the constrained satellite fraction, $f_{\rm sat} \simeq 0.3$, which is about $50\%$ higher than the value reported by \citet{Thornton} using galaxy--galaxy lensing of photometrically selected dwarfs. \citep{Thornton} noted that color cuts in sample selection can bias the inferred $f_{\rm sat}$, making their value a lower bound. In contrast, our sample selection is based on stellar mass and redshift, avoiding this bias. \citet{ZuandMandelbaum} constrained $f_{\rm sat}$ from galaxy--galaxy lensing and clustering of spectroscopic samples in a mass range overlapping our high-stellar-mass bin, finding results consistent with ours within $1\sigma$. Finally, \citet{group08} measure $f_{\rm sat}$ of galaxies with  $\log{M_*}/{\Msun}=9.2$ using the group catalog derived from SDSS \citep{Yang07} and find consistent results with ours.

Given the posterior of the stellar mass--halo mass relation, we populate the IllustrisTNG-300 Dark simulation and compute $f_{\rm sat}$ after applying stellar mass cuts. The derived values are predictions from $\LCDM$ simulations\footnote{We note that our satellite definition is different from those in the literature. We identify the Splashback subhalos and assign those as satellites, as detailed in section \ref{sec:model}, while most of the literature assigns those as isolated halos \citep[e.g.][and references therein]{risaawesomepaper}. As a result, the satellite fractions are higher than the value in the literature. However, we note that we are self-consistent in this paper on the definition of satellites when fitting to the weak lensing profile and generating the $\LCDM$ predicted value.} and are further compared to the measurement in Figure~\ref{fig:fsat}. The predicted value is $\sim 1.5\sigma$ higher than our constraint. We investigate three possible explanations for this discrepancy below. We note that while we explore model variants below, the stellar mass--halo mass relation presented in section~\ref{sec:smhm} is barely affected by those modeling choices (see Fig.~\ref{fig:contour} for a comparison.). 

\subsubsection{Halo properties}

The differences in $f_{\rm{sat}}$ may arise from the choice of halo property used to match $M_*$ in the galaxy–halo model. As shown in \citep{reddick2013}, different choices lead to different satellite fractions. We therefore test three additional halo property variants to match $M_*$: 
\begin{enumerate}
    \item $v_{\rm{peak}}$: peak maximum circular velocity over the halo formation histories. 
    \item $v_{\rm{vir}} $ @ $M_{\rm {peak}}$: virial velociy of the halo at the epoch of $M_{\rm{peak}}$.
    \item $v_{\rm{\alpha}}$: as defined in \citep{lehmann}, $v_\alpha = v_{\rm{vir}} \left(\frac{v_{\rm{max}}}{v_{\rm{vir}}}\right)^{0.57}$, where $v_{\rm{vir}}$ is the virial velocity at the epoch of $M_{\rm{peak}}$ and $v_{\rm{max}}$ is the maximum circular velocity evaluated at the epoch of $M_{\rm{peak}}$. 
\end{enumerate}

In each case, we repeat our analyses to obtain new posteriors of the stellar mass--halo property relation. 
Consitent with \citep{lehmann, reddick2013}, the $\LCDM$-predicted  $f_{\rm{sat}}$ is higher than our fiducial analysis, which adopts $\rm{M}_{\rm{peak}}$ as the halo property. However, we find that in either case, our considered data vector, the galaxy--galaxy lensing and stellar mass function, prefers a lower $f_{\rm{sat}}$, increasing the discrepancy to $\sim 2.5\sigma$. 

\subsubsection{Orphan subhalos}

The subhalos in dark-matter-only simulations can be physically and unphysically disrupted \citep{Klypin1999, 2008ApJ...678....6W, 2018MNRAS.475.4066V, 2024ApJ...970..178M}. Without accounting for this disruption, our estimations of the satellite fractions can be biased. We account for the physical and unphysical disruptions of subhalos using a prescription similar to \cite{universemachine}. Specifically, in each snapshot of the simulation, we find subhalos that are no longer detected by \textsc{Rockstar} in the subsequent snapshots. 
We then locate the host halos that contain these subhalos within their virial radius and use the semianalytic model \cite{universemachine} to evolve the subhalos' position, mass, and maximal circular velocity.  Some of these subhalos will be tidally disrupted when falling into big halos. To determine which of these added subhalos are disrupted, we compare their current maximum circular velocity ($v_{\rm{max}, now}$) to the value at the time when the halo reached its peak mass ($v_{\rm{max}, M_{\rm{peak}}}$). A significantly reduced $v_{\rm{max}, now}$ compared to $v_{\rm{max}, M_{\rm{peak}}}$ indicates that the subhalo has likely undergone substantial tidal stripping and is less likely to host a galaxy. To account for this, we introduce a threshold on the ratio $v_{\rm{max}, now} / v_{\rm{max}, M_{\rm{peak}}}$, below which subhalos are considered too disrupted to host galaxies. 

Because the susceptibility to disruption depends on galaxy properties—such as luminosity, which affects both dynamical friction timescales and resistance to tidal forces—we allow this disruption threshold to vary with $v_{\rm{max}, M_{\rm{peak}}}$. Following \citep{universemachine, cardinal}, we model the probability that a subhalo is physically disrupted as:

\begin{equation}
\label{eq:SHAM_disrupt1}
P(\rm{disrupt}) = \Theta\left(T_{{\rm{dis}}}(v_{\rm{max}, M_{\rm{peak}}})-\frac{v_{\rm{max}, now}}{v_{\rm{max}, M_{\rm{peak}}}}\right), 
\end{equation}
\begin{align}
&T_{{\rm{dis}}}(v_{\rm{max}, M_{\rm{peak}}})= T_l+(T_h-T_l) \nonumber\\
&\times \left( 0.5+0.5{{\rm{erf}}}\left(\frac{\log_{10} v_{\rm{max}, M_{\rm{peak}}}-2.75}{0.25\sqrt{2}}\right)\right),
\label{eq:SHAM_disrupt2}
\end{align}
where $\Theta$ is the Heaviside step function and $\rm{erf}$ is the error function. The function $T_{{\rm{dis}}}(v_{\rm{max}, M_{\rm{peak}}})$ smoothly interpolates between two asymptotic thresholds: $T_l$ at low $v_{\rm{max}, M_{\rm{peak}}}$ and $T_h$ at high $v_{\rm{max}, M_{\rm{peak}}}$. The $T_l$ and $T_h$ values are taken from \citep{saga5} and \citep{universemachine}, respectively. We find around $\simeq 14\%$ of the satellites at $M_{\rm{peak}}=10^{11}\Msun$ are orphan subhalos.

We then repeat the analysis with these added subhalos. As expected, the $\LCDM$ satellite fractions go up when including orphan subhalos. However, because orphan subhalos produce weaker lensing signals, the relative difference in the two-halo terms between satellites and centrals becomes smaller. When fitting the same weak-lensing profile, this reduced contrast leads to a lower inferred $f_{\rm sat}$, which falls $2\sigma$ below the $\Lambda$CDM prediction.

\subsubsection{Baryonic effects}
We further test whether baryonic effects can change the satellite fraction in $\LCDM$ using the IllustrisTNG-300 hydro simulations \citep{TNG}. We select subhalos that are within the stellar mass bins in our analysis. To mimic our satellite definitions, we find the maximum halo mass over the formation histories $M_{\rm{peak}}$ for each subhalo using the SubLink tree \citep{sublink}. We then define a galaxy as a satellite with the same criteria described in Section \ref{sec:model}. We find that the satellite fraction is around $0.6$, slightly higher than our model based on the dark matter only simulation. To test the impact of resolution, we repeat the measurement on IllustrisTNG-100, which has $\sim10$ times higher resolution than IllustrisTNG-300. We find a similar conclusion.\\

The three tested possibilities, including halo properties, orphan subhalos, and baryonic effects, are unlikely to explain the hint of discrepancy between the fiducial inferred $f_{\rm{sat}}$ and the $\LCDM$ predictions (Fig.~\ref{fig:fsat}). We further assess the effect of IIP weights.  As shown in Figure~\ref{fig:weight}, the IIP weights primarily impact the large-scale excess surface matter density profile, where the $f_{\rm{sat}}$ information comes from (see Figure 10 of \citep{Thornton}). We find a factor of two difference between the inferred $f_{\rm{sat}}$ with and without IIP weights. This indicates the importance of IIP weights to the inferred $f_{\rm{sat}}$ value. \citep{desilensing} shows that the IIP weights can remove the impact of fiber incompleteness to below $1\%$, which is $20$ times smaller than the statistical uncertainties of this work. 
To further test the sensitivity of our analysis to the accuracy of the IIP weight\footnote{In appendix \ref{sec:alternativeweight}, we test the impact of an alternative IIP weight and find that it does not affect the conclusion of this paper.}, we artificially change $\Delta \Sigma$ by $1\%$ and $3\%$ and find a shift of $f_{\rm{sat}}<0.024\sigma$ and $f_{\rm{sat}}<0.14\sigma$ respectively, where $\sigma$ is the statistical uncertainty. In appendix \ref{sec:fiberassignment}, we further perform dedicated simulations to quantify the validity of IIP weights on removing the fiber incompleteness problem for dwarf galaxies. In principle, IIP weights should be sufficiently accurate to remove the impact of fiber incompleteness. However, since IIP weights remove the dominant systematics in this measurement, advanced simulations that fully resemble the DESI survey might be needed to quantify the accuracy of the IIP weights for dwarf samples. Since the discrepancy between the inferred $f_{\rm{sat}}$ and $\LCDM$ predictions is not statistically significant, we leave these investigations to future work. 

\section{Conclusions}
\label{sec:conc}
Leveraging the large overlap between the DESI-DR1 spectroscopic survey and the DECADE/DES weak lensing survey data, we present the first detection of weak gravitational lensing profiles around spectroscopically confirmed dwarf galaxy samples ($\log(M_*/\Msun)<10^{9.2}$). To avoid complications in modeling selection functions, we select clean dwarf galaxy samples out of the DESI-DR1 BGS data with redshift and stellar mass cuts.
Employing a simulation-based modeling approach, we extract the stellar mass--halo mass relation and the satellite fraction of the sample from the measurement.  Our main findings are: 
\begin{enumerate}
    \item Using DECADE and DES weak lensing data, we measure the excess surface mass density profiles of the selected low mass DESI galaxies with stellar mass in $\log \rm{M}_*=[8.2, 9.2] \Msun$ and $\log \rm{M}_*=[9.2, 10.2] \Msun$ and find a signal to noise of $5.6$ and $12.4$. We find this measurement is robust against different stellar mass estimates, boost factors, weak lensing calibration systematics, and the photometry of shredded objects.  
    \item We find that the incompleteness of fiber assignments is a significant systematic in the lensing measurements on large scales and correct it using the individual inverse probability (IIP) weight \citep{DESItarget, desilensing}.
    \item Using a simulation-based modeling approach, we extract the stellar mass -- halo mass relations and the satellite fraction from the measurements. We find that the stellar mass -- halo mass relation is consistent with measurements in the literature, but our measurements indicate somewhat higher halo masses in the considered stellar mass bins than some analytic models.
    
    \item We find a $1.5\sigma$ hint that the constrained satellite fraction is lower than the $\LCDM$ predictions. We test three variants of the analysis by changing the halo properties used to match stellar mass, considering orphan subhalos, and including baryonic effects. We find that none of these alternatives reduce the $f_{\rm{sat}}$ inconsistency hints. Finally, we find that applying the IIP weights can change the constrained $f_{\rm{sat}}$ by a factor of $2$, indicating the importance of the IIP weight accuracy to the $f_{\rm{sat}}$ constraints. We therefore perform dedicated simulation tests in Appendix \ref{sec:fiberassignment} to validate the accuracy of IIP weights for dwarf galaxies.
\end{enumerate}

We note that at the same time as this analysis, Treiber et al. (2025) have carried out an independent analysis using an extended DESI dwarf sample, with a different lensing catalog. The focus of that paper is to assess the consistency of the spectroscopic measurements with those from a photometric dwarf sample built by leveraging the spectroscopy, which results in a higher signal-to-noise measurement. The two analyses are complementary, and we leave a detailed comparison to future work. 

This work provides the first attempt at measuring weak gravitational lensing profiles around spectroscopically confirmed dwarf galaxies, opening a new observational window into the galaxy--halo connection at low masses. With the expanded DESI–DECADE/DES/Rubin overlap, future data releases will deliver significantly higher signal-to-noise measurements, enabling more precise constraints on both the stellar mass–halo mass relation and the satellite fraction. Realizing the full potential of these datasets will require a reliable fiber incompleteness correction scheme, particularly the accuracy of IIP weights for dwarf galaxy samples, to ensure unbiased results. Looking further ahead, using the kinematic lensing technique \citep{2025MNRAS.540.2877P,KL1}, one could increase the lensing constraining power of low-redshift galaxy samples, which would further boost the signal-to-noise ratio. Together, these advances will pave the way for high-precision, robust tests of galaxy formation models and $\LCDM$ predictions in the low-mass regime.

\section{Acknowledgements}
CHT is supported by the Eric and Wendy Schmidt AI in Science Postdoctoral Fellowship, a Schmidt Futures program. CHT thanks the organizers of Understanding cosmological observations at the Benasque Physics Center in 2025, where a significant amount of work was performed. This work was completed in part with resources provided by the University of Chicago’s Research Computing Center. CHT thanks ChangHoon Hahn and Ashley Ross for valuable discussions on BGS target selection, and Tomomi Sunayama and Ashley Ross for discussions on IIP weights. CHT thanks Susmita Adhikari and Annika Peter for providing helpful comments on the draft. We thank Viraj Manwadkar for helpful discussions about stellar mass estimates and photometric shredding. 
CHT thanks Andrey Kravstov for sharing data from  Kravtsov et al. 2022. The DECADE project is supported by NSF AST-2108168 and
AST-2108169. The DELVE Survey gratefully acknowledges support from Fermilab LDRD (L2019.011), the NASA Fermi Guest Investigator Program Cycle 9 (No. 91201), and the NSF (AST-2108168, AST-2108169, AST-2307126, AST2407526, AST-2407527, AST-2407528). 

This project used public archival data from the Dark Energy Survey (DES). Funding for the DES Projects has been provided by the U.S. Department of Energy, the U.S. National Science Foundation, the Ministry of Science and Education of Spain, the Science and Technology Facilities Council of the United Kingdom, the Higher Education Funding Council for England, the National Center for Supercomputing Applications at the University of Illinois at Urbana–Champaign, the Kavli Institute of Cosmological Physics at the University of Chicago, the Center for Cosmology and Astro-Particle Physics at the Ohio State University, the Mitchell Institute for Fundamental Physics and Astronomy at Texas A$\&$M University, Financiadora de Estudos e Projetos, Fundação Carlos Chagas Filho de Amparo à Pesquisa do Estado do Rio de Janeiro, Conselho Nacional de Desenvolvimento Científico e Tecnológico and the Ministério da Ciência, Tecnologia e Inovação, the Deutsche Forschungsgemeinschaft and the Collaborating Institutions in the Dark Energy Survey.

The Collaborating Institutions are Argonne National Laboratory, the University of California at Santa Cruz, the University of Cambridge, Centro de Investigaciones Enérgeticas, Medioambientales y Tecnológicas–Madrid, the University of Chicago, University College London, the DES-Brazil Consortium, the University of Edinburgh, the Eidgenössische Technische Hochschule (ETH) Zürich, Fermi National Accelerator Laboratory, the University of Illinois at Urbana-Champaign, the Institut de Ciències de l’Espai (IEEC/CSIC), the Institut de Física d’Altes Energies, Lawrence Berkeley National Laboratory, the Ludwig-Maximilians Universität München and the associated Excellence Cluster Universe, the University of Michigan, the National Optical Astronomy Observatory, the University of Nottingham, The Ohio State University, the OzDES Membership Consortium, the University of Pennsylvania, the University of Portsmouth, SLAC National Accelerator Laboratory, Stanford University, the University of Sussex, and Texas A$\&$M University.

Based in part on observations at Cerro Tololo Inter-American Observatory, National Optical Astronomy Observatory, which is operated by the Association of Universities for Research in Astronomy (AURA) under a cooperative agreement with the National Science Foundation.

This research used data obtained with the Dark Energy Spectroscopic Instrument (DESI). DESI construction and operations is managed by the Lawrence Berkeley National Laboratory. This material is based upon work supported by the U.S. Department of Energy, Office of Science, Office of High-Energy Physics, under Contract No. DE–AC02–05CH11231, and by the National Energy Research Scientific Computing Center, a DOE Office of Science User Facility under the same contract. Additional support for DESI was provided by the U.S. National Science Foundation (NSF), Division of Astronomical Sciences under Contract No. AST-0950945 to the NSF’s National Optical-Infrared Astronomy Research Laboratory; the Science and Technology Facilities Council of the United Kingdom; the Gordon and Betty Moore Foundation; the Heising-Simons Foundation; the French Alternative Energies and Atomic Energy Commission (CEA); the National Council of Humanities, Science and Technology of Mexico (CONAHCYT); the Ministry of Science and Innovation of Spain (MICINN), and by the DESI Member Institutions: www.desi.lbl.gov/collaborating-institutions. The DESI collaboration is honored to be permitted to conduct scientific research on I’oligam Du’ag (Kitt Peak), a mountain with particular significance to the Tohono O’odham Nation. Any opinions, findings, and conclusions or recommendations expressed in this material are those of the author(s) and do not necessarily reflect the views of the U.S. National Science Foundation, the U.S. Department of Energy, or any of the listed funding agencies.

\bibliographystyle{apsrev}
\bibliography{sample}
\clearpage
\onecolumngrid
\appendix

\section{Comparison of stellar masses}
\label{sec:stellar_mass}
\begin{figure}
\includegraphics[width=0.5\linewidth]{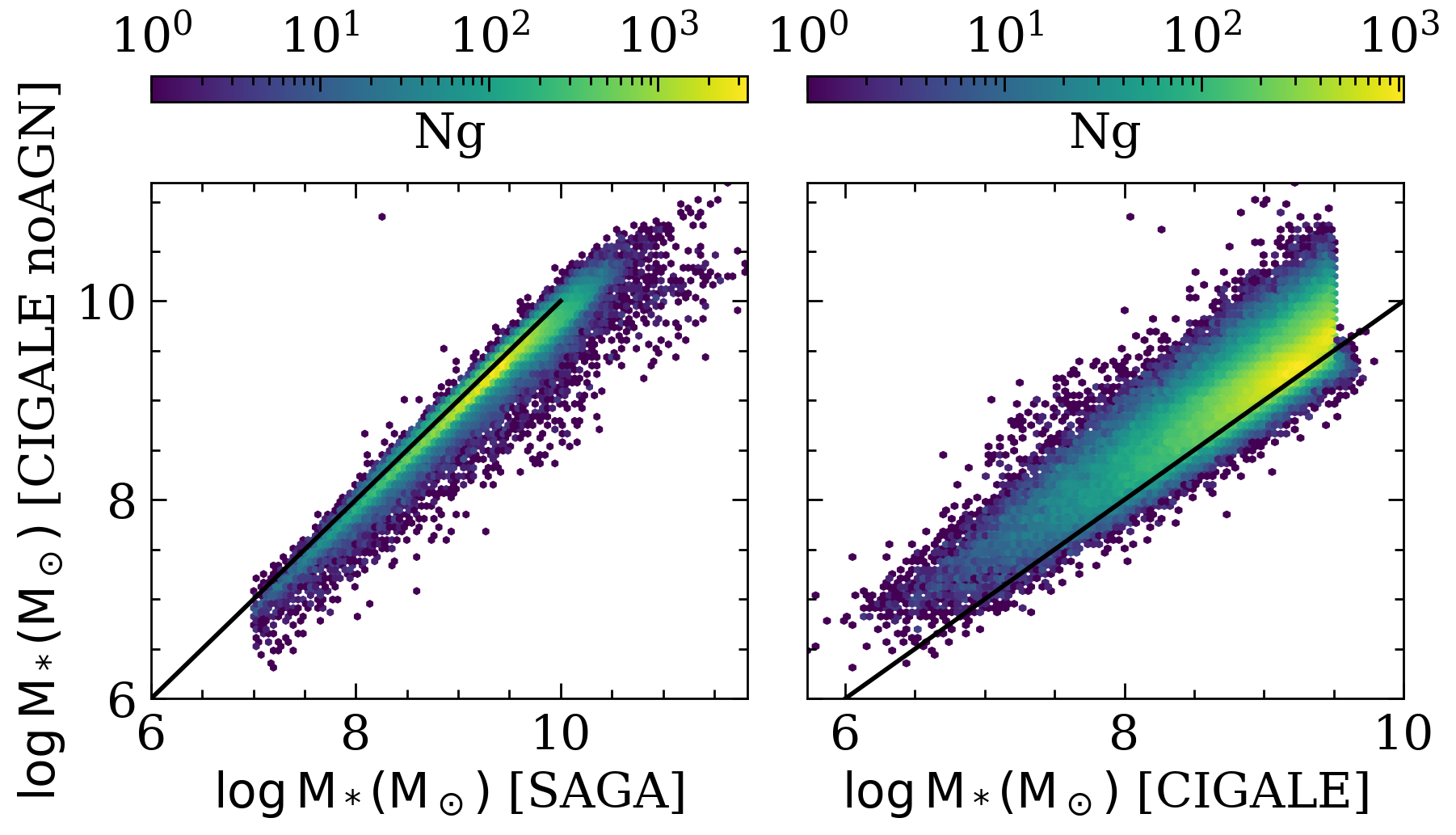}
\caption{Comparisons of stellar mass estimates from CIGALE-noAGN, CIGALE, and SAGA. Color maps indicate the number of galaxies. Black lines show a one-to-one relation.}
\label{fig:stellarmasscom}
\end{figure}

\begin{figure}
\includegraphics[width=0.5\linewidth]{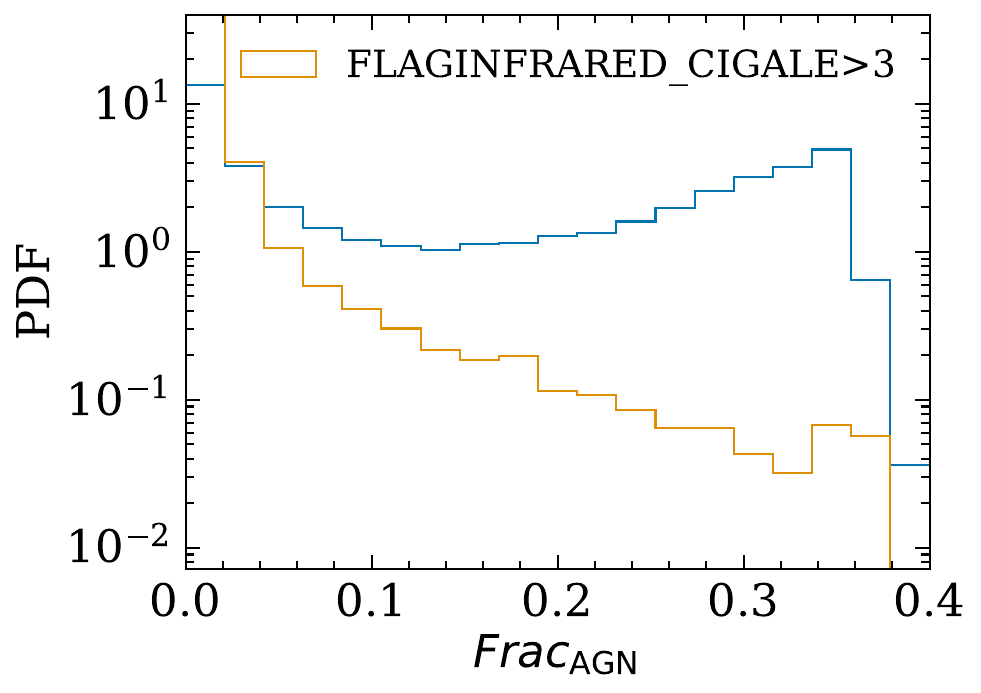}
\caption{Distribution of CIGALE-estimated fractional infrared flux from AGN. The blue line shows the full samples used in this analysis, while the orange line shows the subset of galaxies with more than three infrared bands having signal-to-noise greater than three.}
\label{fig:agnfrac}
\end{figure}

 Figure~\ref{fig:stellarmasscom} compares different stellar mass estimates. We find that the stellar mass from SAGA and CIGALE noAGN shows great agreement, while the CIGALE noAGN and CIGALE show a $0.15$ dex systematic bias. To investigate the origin of this discrepancy, we investigate the robustness of the CIGALE inferred infrared flux fraction from AGNs. Figure~\ref{fig:agnfrac} shows the distribution of CIGALE-inferred fraction of infrared flux from AGN (\texttt{FracAGN}). Since this estimation highly depends on the quality of infrared data, we further compare the same distribution with subsets of galaxies that have at least three infrared bands with signal-to-noise greater than three. As indicated in \citep{CIGALE}, the AGN fraction is more reliably constrained when high signal-to-noise infrared is available. Figure~\ref{fig:agnfrac}  shows that the median value of the full sample's \texttt{FracAGN} is biased high by $\simeq0.15$ dex compared to the samples with high-quality infrared data. This is consistent with the expectation that CIGALE mass is biased low compared to CIGALE-noAGN and SAGA. Detailed correction of CIGALE is beyond the scope of this work.  

\section{Lensing systematics}
\label{sec:alllensingsys}
In this section, we test the impact of various lensing systematics to the measurement. 
\subsection{Stellar mass estimates}
\begin{figure}
\includegraphics[width=\linewidth]{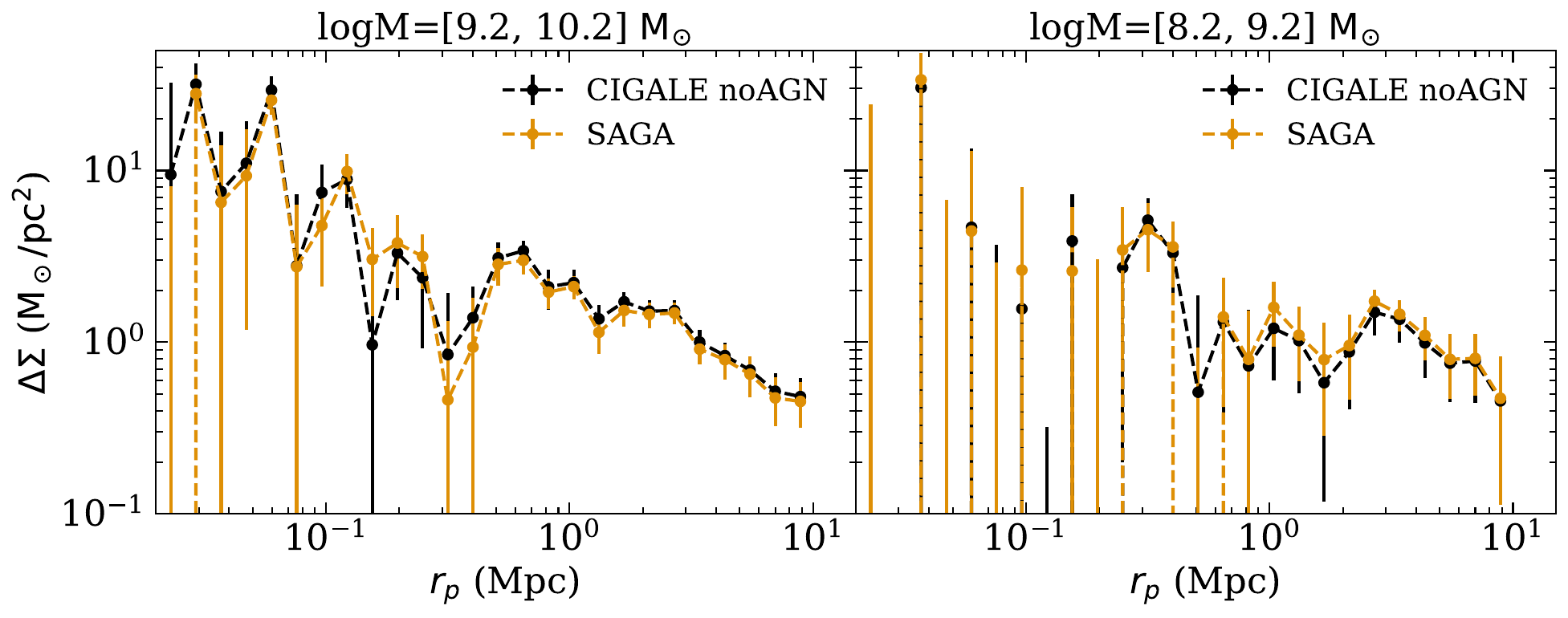}
\caption{Diagnostic plot for the lensing measurements of galaxies in two stellar mass bins. Error bars show $1\sigma$ uncertainties. Black dots correspond to the fiducial analysis, while the brown dots show alternative stellar mass estimates.}
\label{fig:cigale_saga}
\end{figure}

\begin{figure}
\includegraphics[width=\linewidth]{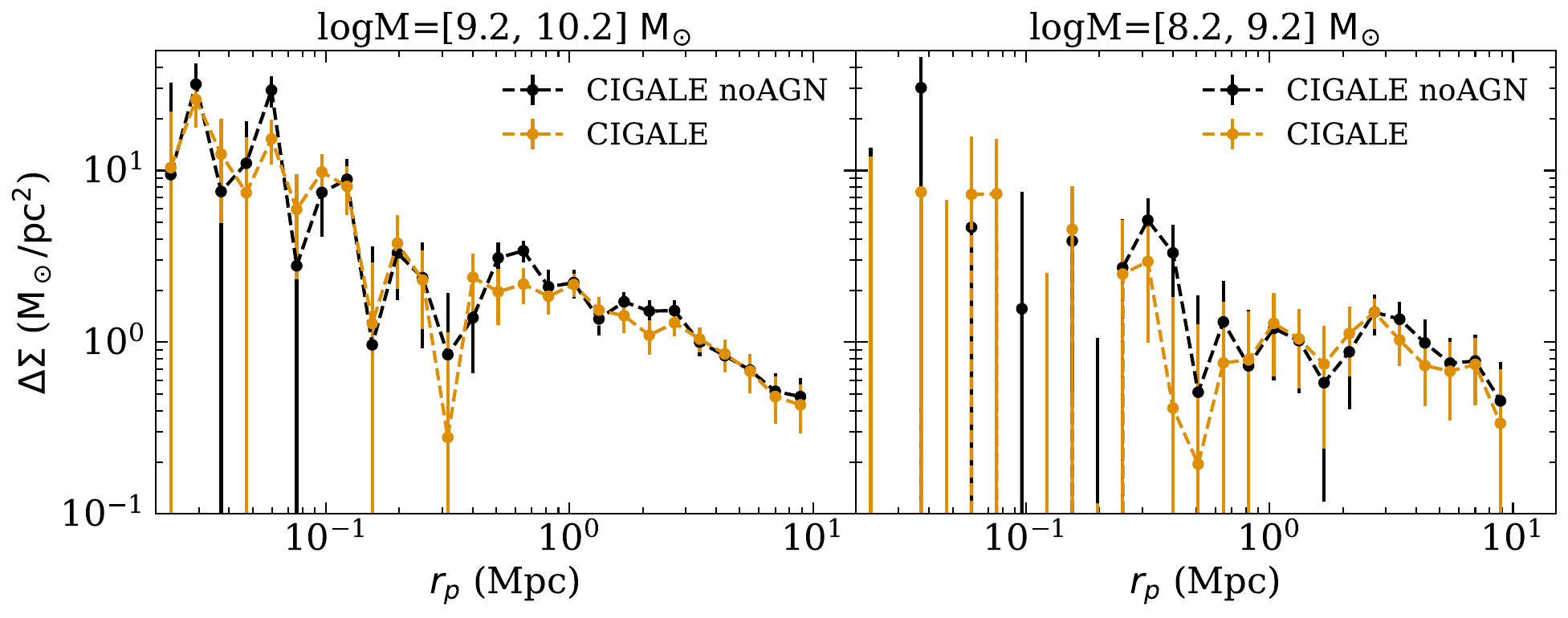}
\caption{Same as Fig.~\ref{fig:cigale_saga} but the brown dots correspond to measurements with CIGALE stellar estimations.}
\label{fig:cigale_cigalenoagn}
\end{figure}

Figure \ref{fig:cigale_saga} compares the $\Delta \Sigma$ using CIGALE-noAGN and SAGA stellar mass estimations. To ensure an apples-to-apples comparison, we perform this analysis in an abundance matching way. Specifically, we define the stellar mass cut in SAGA stellar mass so that the number density of dwarf galaxies above the cut is the same as the CIGALE-noAGN corresponding stellar mass cut. Comparing the two data vectors, we find $\Delta\chi^2~=~10$ with $52$ degrees of freedom, indicating the robustness of the measurement with respect to different stellar mass estimations. Alternatively, we compare $\Delta \Sigma$ using CIGALE-noAGN and CIGALE stellar mass estimations in Fig.~\ref{fig:cigale_cigalenoagn}, we find a much larger difference ($\Delta \chi^2=35$) with the same degrees of freedom. This might indicate our measurement has non-negligible systematics due to uncertainties in stellar mass estimation. However, as discussed in Appendix~\ref{sec:stellar_mass}, we find that CIGALE likely underestimates the stellar masses of DESI dwarf galaxies due to an overestimated AGN contribution.


\subsection{Multiplicative shear biases}
\begin{figure}
\includegraphics[width=\linewidth]{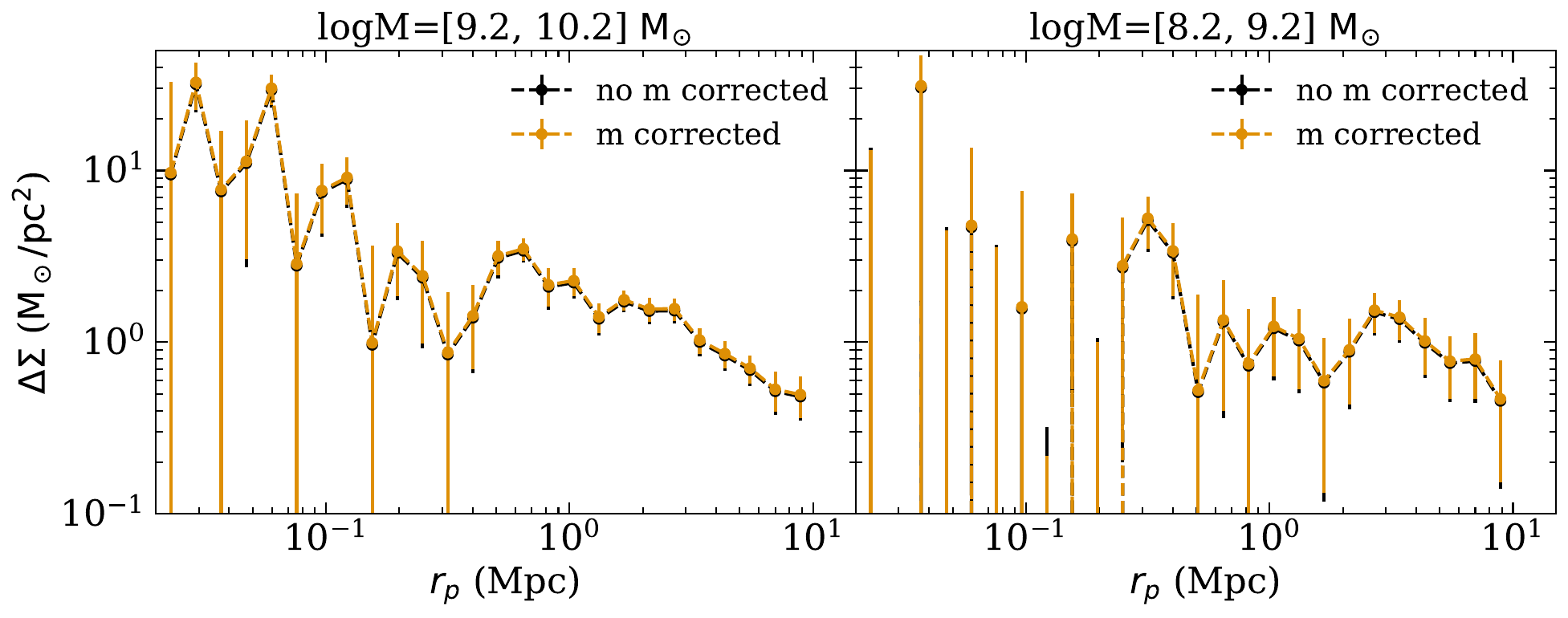}
\caption{Same as Fig.~\ref{fig:cigale_saga} but the brown dots correspond to measurements with multiplicative bias corrections.}
\label{fig:cigale_mom}
\end{figure}

Weak lensing measurement can suffer from additive and multiplicative biases. While additive biases can be mitigated by subtracting the mean of the shears in the catalogs, the multiplicative biases must be explicitly calibrated. A lensing survey like DES and DECADE relies on imaging simulations to quantify this bias. Fig.~\ref{fig:cigale_mom} shows a comparison of $\Delta \Sigma$ with and without applying the best-fit multiplicative shear biases from \citep{y3-imagesims, Anbajagane2025a}. The $\Delta\chi^2$ between the two measurements is $0.14$, indicating that such a bias is negligible given the current precision of the measurement. 
\subsection{Shredded objects}
\begin{figure}
\includegraphics[width=\linewidth]{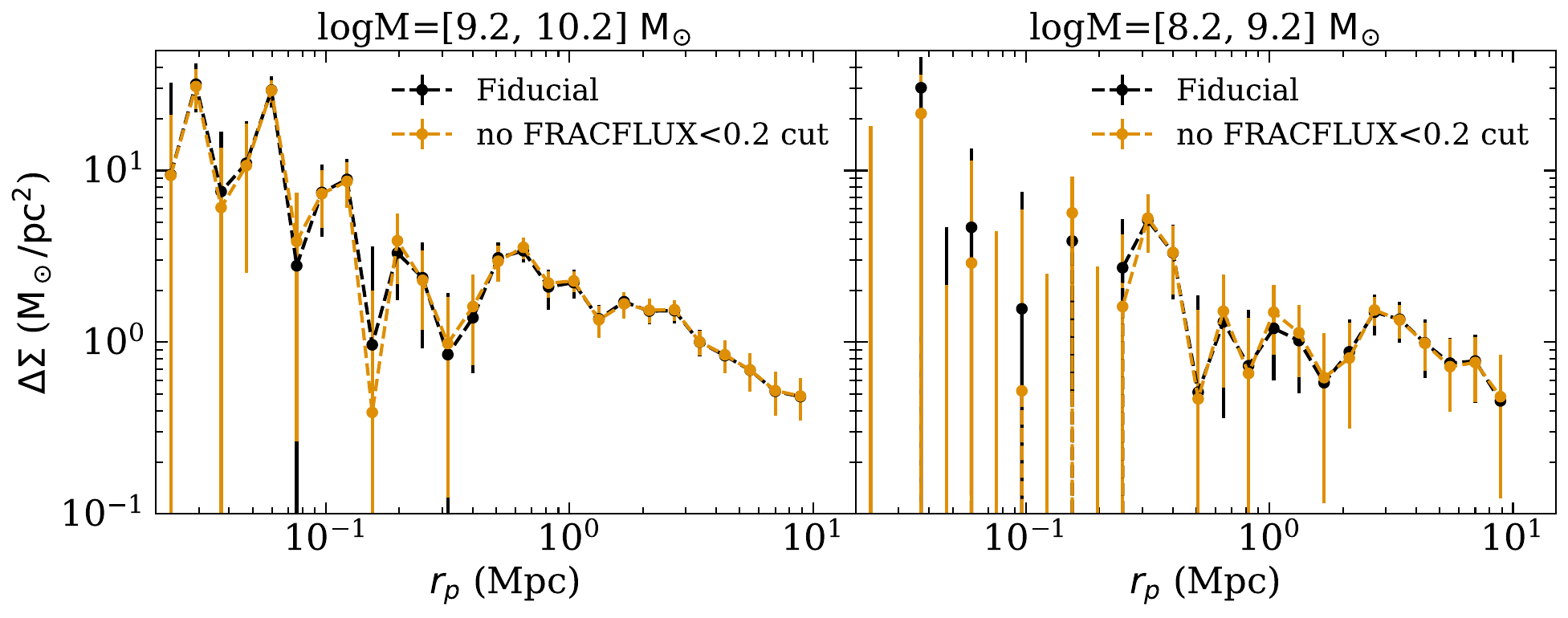}
\caption{Same as Fig.~\ref{fig:cigale_saga} but the brown dots correspond to measurements without \texttt{FRACFLUX} cut.}
\label{fig:fracvut}
\end{figure}

Dwarf galaxies can be shredded into multiple objects \citep{Elise, 2023ApJS..269....3M,2012AJ....143..133H}. In the fiducial analysis, we apply \texttt{FRACFLUX}$<0.2$ in $g,r,z$ bands cut to mitigate this problem. Figure \ref{fig:fracvut} shows a comparison of $\Delta\Sigma$ with and without \texttt{FRACFLUX} cut and finds a $\Delta \chi^2=0.62$. 

\subsection{Boost factor}
\begin{figure}
\includegraphics[width=\linewidth]{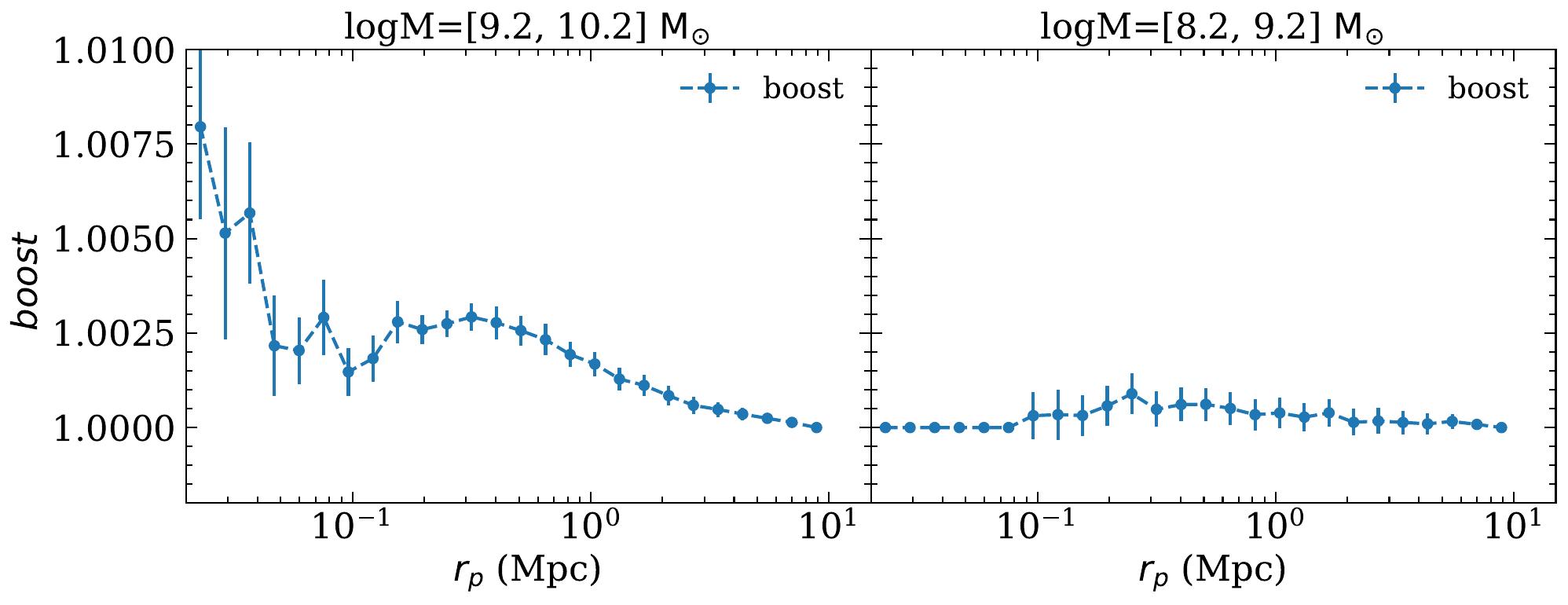}
\caption{Boost factor as a function of scale in two different stellar mass bins. Error bars show $1\sigma$ uncertainties.}
\label{fig:boostnoboost}
\end{figure}

Figure \ref{fig:boostnoboost} shows the measurement of boost factors defined in equation \ref{eq:boost}. As expected, the high stellar mass sample has a larger boost factor due to more signficant redshift overlap with source galaxies (Fig.~\ref{fig:nz}). 
The boost factor is much smaller than measurements around photometrically selected dwarf samples \citep{Thornton}, indicating the effectiveness of spectroscopic redshifts in separating the redshift range of lens and source galaxies.

\subsection{Lensing B mode}
\begin{figure}
\includegraphics[width=0.7\linewidth]{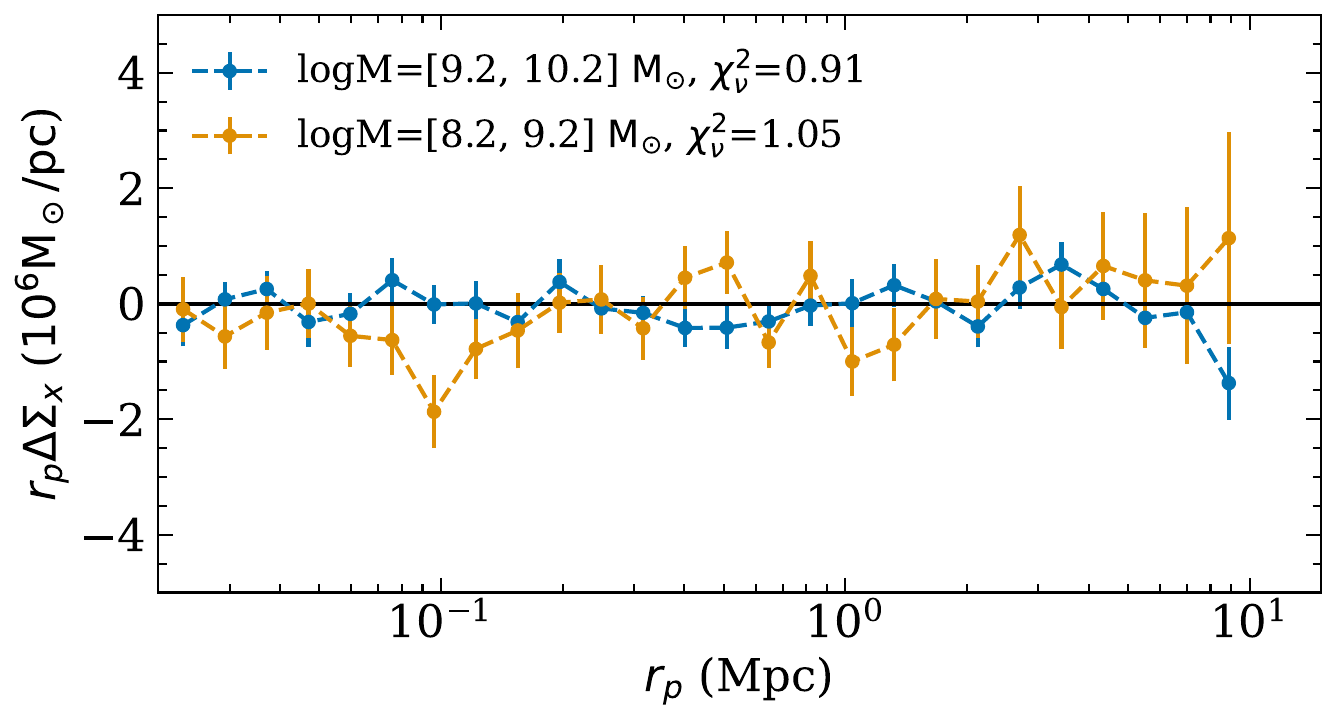}
\caption{ Null test of the lensing measurement. Cross-component of the weak lensing profile (or B-mode components of shears) around DESI galaxies in two stellar mass bins. Error bars show $1\sigma$ uncertainties.}
\label{fig:bmode}
\end{figure}

The B-mode components of shears around lens galaxies are expected to be zero due to parity invariance. The measurement of the B-mode component, therefore, serves as a null test of the measurement. Figure \ref{fig:bmode}  shows the measurement of the B-mode components for dwarf galaxies in two stellar mass bins, and finds the measurements consistent with the null hypothesis. 
\section{Parameters constraints} 
\begin{figure}
\includegraphics[width=\linewidth]{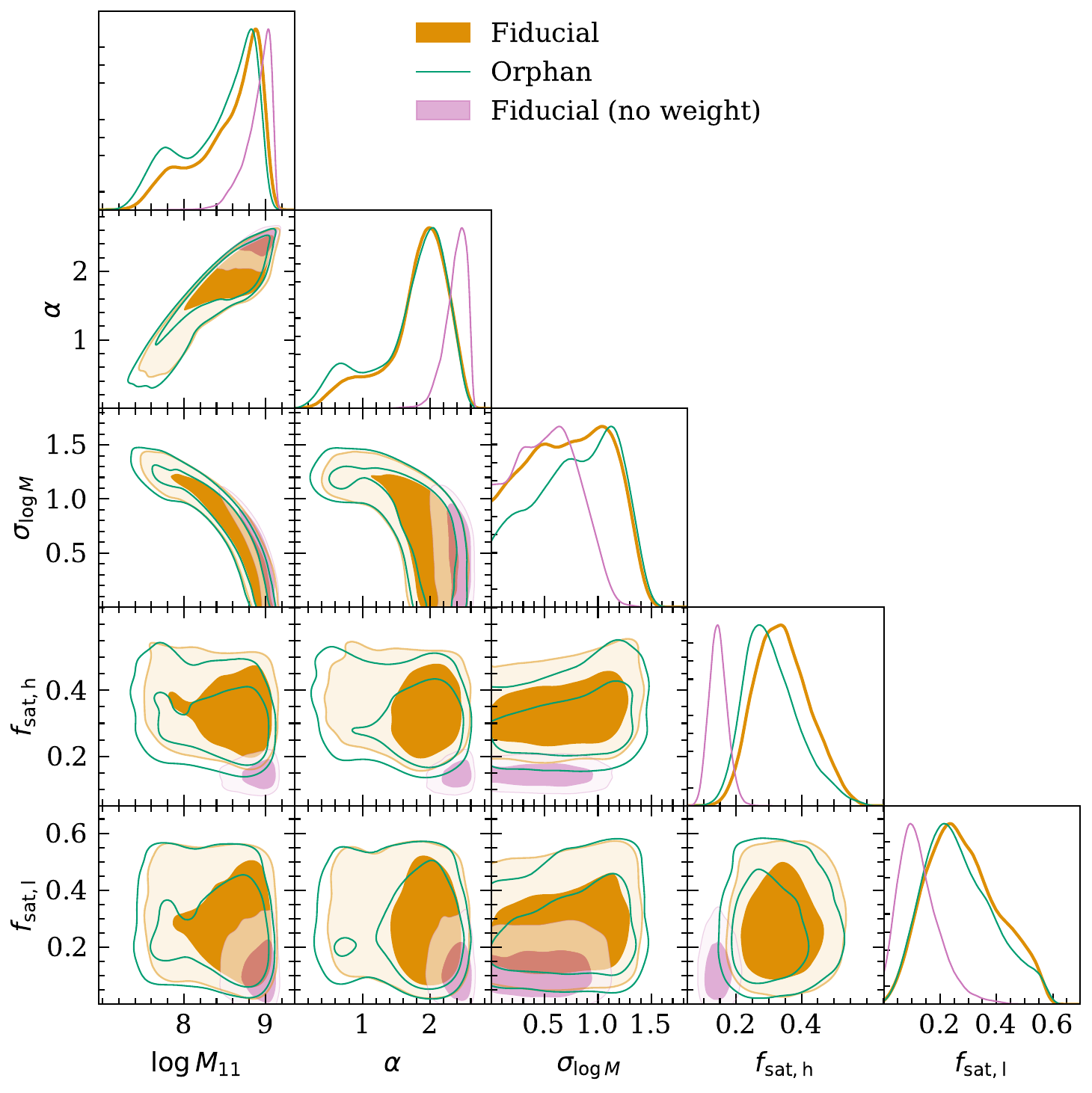}
\caption{Full posteriors for the model fits. Contours show $68\%$ and $95\%$ credible intervals. Brown contours correspond to the fiducial analysis, green contours correspond to the analysis including orphan subhalos, and the pink contours correspond to the analysis on data vectors generated without applying IIP weights.}
\label{fig:contour}
\end{figure}

\begin{figure*}
\includegraphics[width=0.6\linewidth]{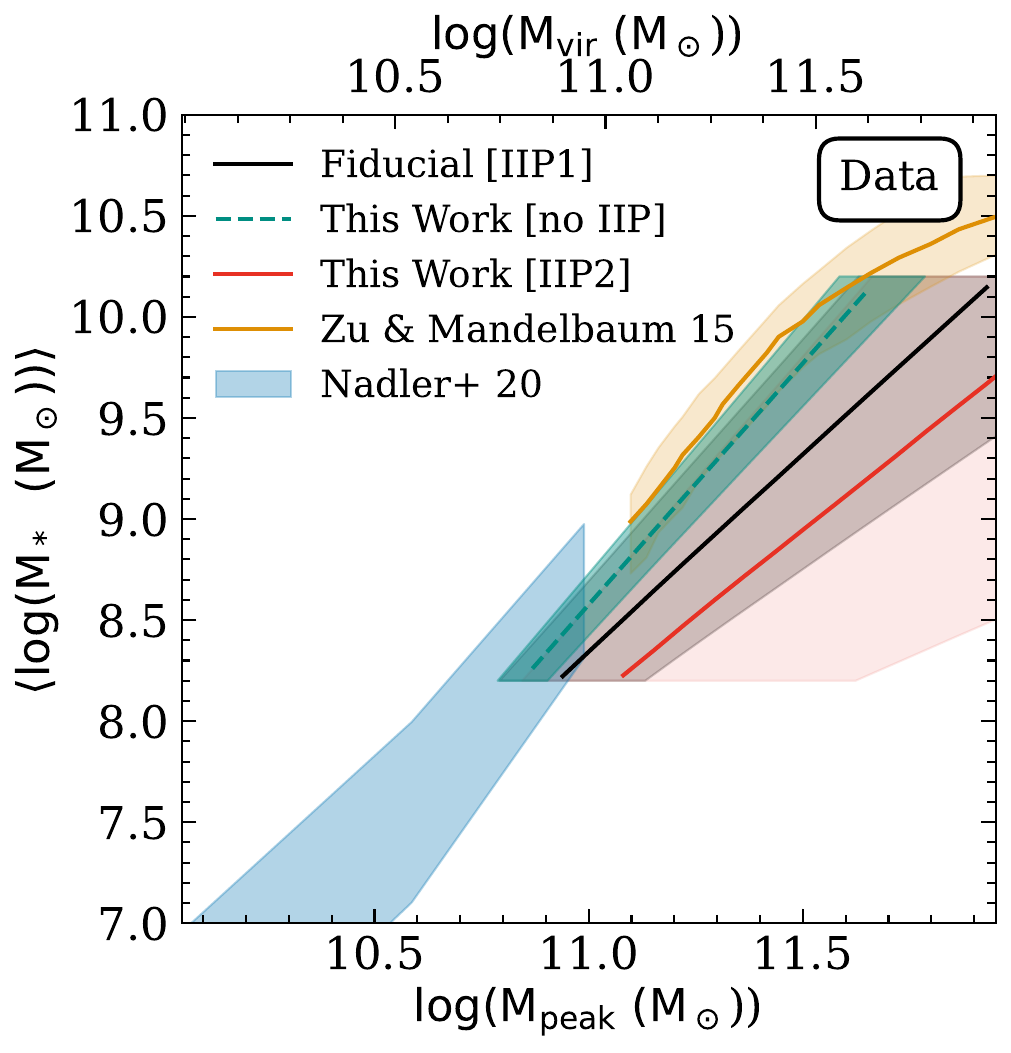}
\caption{
Impacts of weights on the inferred stellar mass--halo mass relation. The black line shows the best-fit value of the fiducial analyses, and the shaded black regions show the $68\%$ credible intervals. The green line and shaded regions show the best-fit and credible regions obtained from analyses of measurements without IIP weights.  The red line and shaded regions show the best-fit and credible regions obtained from analyses of measurements with alternative IIP weights. Similar to Fig.~\ref{fig:smhm}, we overlay the constraints from different literatures. The brown line shows constraints from the joint analysis of galaxy--galaxy lensing and galaxy clustering in SDSS \citep{ZuandMandelbaum} with shaded regions showing $1\sigma$ uncertainties. The blue shaded region shows constraints from Milky Way satellite abundances \citep{Nadler21}. 
}
\label{fig:smhm_noiip}
\end{figure*}

Figure \ref{fig:contour} shows the posterior of the simulation-based model sampled by the MCMC chains. The fiducial analysis and the analysis including orphan subhalos are largely consistent, indicating the robustness of the result. On the other hand, we find that the analysis on the measurement without IIP weights is significantly different from the fiducial analysis. To illustrate this difference, we overplot the best-fit stellar mass--halo mass relation from the analysis on the measurement without IIP weights in Fig.~\ref{fig:smhm_noiip}.  

\begin{figure*}
\includegraphics[width=1\linewidth]{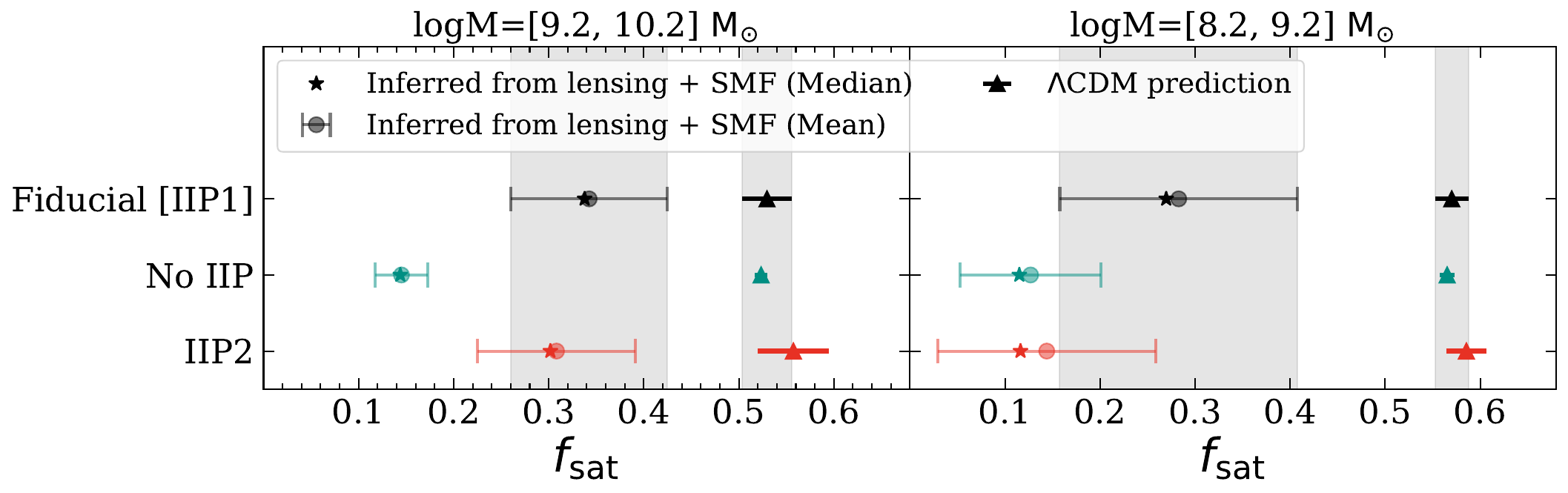}
\caption{
Impacts of weights on the inferred $f_{\rm{sat}}$. Error bars indicate the $68\%$ credible intervals. Dots show the marginalized means from weak lensing profile analyses with stellar mass function priors, while stars show the median of the marginalized constraints. The triangles show the mean $\Lambda$CDM prediction based on the stellar mass–halo mass relation constrained from those same analyses. Each row corresponds to a different analysis variant. The black dots correspond to the fiducial analysis, the green dots correspond to the analysis on the data vector generated without weights, and the red dots correspond to the analysis on the data vector generated with an alternative IIP weight. }
\label{fig:fsat_weight}
\end{figure*}

\section{Alternative weight}
\label{sec:alternativeweight}
Recently, DESI released the v1.5pip catalog, which includes an alternative method for quantifying the weights used to mitigate fiber incompleteness, summarized in \citep{2025JCAP...01..127L}. In this approach, DESI generated 128 additional fiber-assignment realizations and recorded the fraction of times each galaxy is assigned a fiber, denoted \textsc{PROB\_OBS}. The corresponding IIP weight is then defined as $129/(1+128,\rm{PROB\_OBS})$ \citep{2018MNRAS.481.2338B}. For clarity, we refer to this weight as IIP2, while the weights used in our fiducial analysis are denoted IIP1. Since the DESI galaxy–galaxy lensing measurements \citep{2025arXiv250621677H} adopt IIP1, we also use IIP1 for our fiducial results, and here test the sensitivity of our conclusions to the choice between IIP1 and IIP2.

We find that measurements using IIP2 are substantially noisier than those with IIP1, consistent with Appendix C of \citep{2025JCAP...07..017A}. Specifically, the signal-to-noise drops to $4.6$ and $6.9$ in the two stellar mass bins ($\log \rm{M}_*=[8.2, 9.2]~\Msun$ and $\log \rm{M}_*=[9.2, 10.2]\Msun$), respectively. Fitting the model to the IIP2-based measurements, we obtain the stellar mass–halo mass relation and satellite fraction. The stellar mass–halo mass relation is nearly unconstrained (Fig.\ref{fig:smhm_noiip}), with lower stellar masses inferred at fixed halo mass, likely due to the larger prior volume at low stellar mass. The inferred $f_{\rm sat}$ is shifted to even lower values (Fig.~\ref{fig:fsat_weight}). We therefore conclude that there is no evidence that adopting IIP2 affects our fiducial results.
\section{Fiber assignment simulation}
\label{sec:fiberassignment}
We use simulations to test the validity of IIP weights for mitigating fiber incompleteness systematics. While a similar test was performed in \citep{desilensing}, there are two key differences that make its applicability to our work uncertain. First, \citep{desilensing} tested the full BGS sample, whereas our analysis focuses on a subset of BGS galaxies. Second, their study examined IIP2 weights, which are derived from multiple realizations of fiber assignments, while our main analysis uses IIP1 weights, constructed from the number of objects competing for a fiber in each observation.

To address these differences, we generate a simulation using the best-fit model from our analysis and use it to evaluate the performance of the IIP weights. Specifically, we assign stellar masses to halos following Eq.~\ref{eq:stellarmasshalomass}, then compute $r$-band absolute magnitudes using the SAGA relation \citep{saga3}, assuming $g-r$ colors equal to the median value of our dwarf sample. We build a lightcone from the simulation box with the observer placed at $(0,0,0)$, yielding a volume with $z<0.1$ that covers $8100\ \rm{deg}^2$. Distances are converted to redshifts, which are used to convert absolute magnitudes into apparent magnitudes. Galaxy positions are rotated so that the survey footprint is centered at $\mathrm{RA}=180^\circ$ and $\mathrm{Dec}=30^\circ$.

We then select galaxies with $m_r<19.5$ as BGS bright and $19.5<m_r<20.175$ as BGS faint. Because the catalog is limited to $z<0.1$, the resulting number density is lower than in the real BGS. This will lead to an artificially high completeness given the same observational time. To correct for this, we supplement the catalog with galaxies at random positions such that the number densities match $800\ \mathrm{deg}^{-2}$ for the bright sample and $500\ \mathrm{deg}^{-2}$ for the faint sample. Note that the fiber assignment algorithm only requires galaxies' positions and categories, so we do not need to generate magnitudes for these galaxies. Since galaxies at different redshifts are uncorrelated, this random augmentation is a valid approximation. During the bright time survey, the Milky Way white dwarfs will have higher probability of being assigned a fiber, it is therefore important to include those samples in the simulation. We generate those samples assuming a number density of $1$ per $\mathrm{deg}^{-2}$ at random position. 

We then assign the priority of $2100$, $2000$, and $2988$ to BGS bright galaxies, BGS faint galaxies, and Milky Way white dwarfs respectively. We run the DESI's fiber assignment code\footnote{https://github.com/desihub/fiberassign/tree/main} based on the DESI model\footnote{https://desi.lbl.gov/svn/code/desimodel/trunk/data/}. From the simulated catalog, we generate two BGS bright galaxy samples with $z<0.1$.
First, we include all BGS bright galaxies that are reachable by DESI fibers. This sample is free from fiber assignment incompleteness and thus serves as our reference (or “parent”) sample. Second, we select only the BGS bright galaxies that were actually assigned a fiber, which mimics the observed data. For this observed sample, we construct IIP1 weights by counting the number of BGS bright galaxies competing for a fiber in each observation; the inverse of this quantity corresponds to \textsc{FRACZ\_TILELOCID} in the DESI LSS catalog.

In addition, DESI defines another incompleteness measure, \textsc{FRAC\_TLOBS\_TILES}, which quantifies fiber competition between different galaxy categories. Because BGS bright targets have high priority, the effect of this component is expected to be negligible. Indeed, we find in the data that excluding \textsc{FRAC\_TLOBS\_TILES} in the IIP weight calculation resulting in a change in weak lensing profile at $\Delta\chi^2=0.002$. We therefore ignore this component when constructing IIP weights in simulations.

\begin{figure*}
\includegraphics[width=1\linewidth]{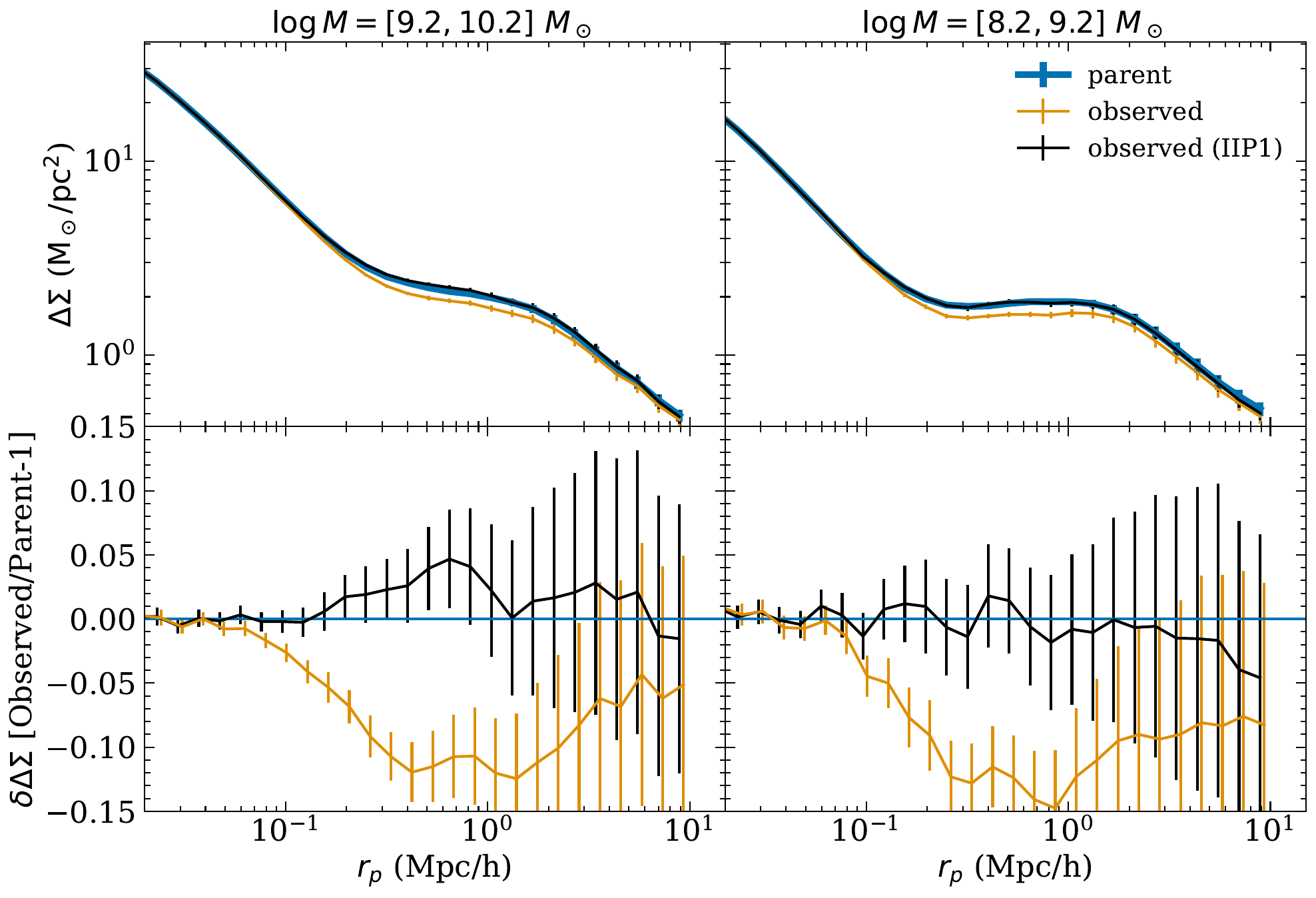}
\caption{
Simulation tests of the validity of IIP weights. Error bars show $1\sigma$ uncertainties estimated with 100 Jackknife resampling. 
Blue lines show the lensing signal around the parent sample, while the orange lines show that of the observed sample. The black line shows the lensing profile around the observed sample weight by IIP1. }
\label{fig:fiber}
\end{figure*}

For both the parent and observed samples, we further select subsets of galaxies based on stellar mass to match the cuts used in the data. We then measure the weak lensing profiles around these samples (Fig.~\ref{fig:fiber}). As expected, fiber incompleteness produces a noticeable impact on the lensing signal, consistent with the findings of \citep{desilensing}. However, when we apply the IIP1 weights to the observed sample, the recovered weak lensing profile matches that of the parent sample. This demonstrates that IIP weights effectively correct for fiber incompleteness in dwarf galaxy lensing measurements. We do not expect the measured difference between the reference and observed weak lensing profiles to precisely match that seen in the data (Fig.~\ref{fig:weight}) as the simulated catalog only covers a part of the data sky footprint, due to computational limitations. This results in different fiber assignment configurations between the simulations and the data. We note that these differences are only stochastic and do not indicate any systematic differences in the simulations compared to the data.

\end{document}